\documentclass[aps,pra,floatfix,showpacs,preprint,superscriptaddress]{revtex4-2}
\usepackage{graphicx}
\usepackage[dvipsnames]{xcolor}
\usepackage{blindtext}
\usepackage{geometry}
\usepackage{amsmath}
\usepackage{amssymb}
\usepackage{slashed}
\usepackage{feynmp-auto}
\usepackage{enumitem}
\usepackage{braket}
\usepackage{bm}
\usepackage{mathrsfs}
\usepackage{revsymb}
\usepackage{accents}
\usepackage{color}
\usepackage{hyperref}
\usepackage[compat=1.1.0]{tikz-feynman}
\usepackage[build={latex=pdflatex}]{standalone}

\newcommand{\F}{{\mathcal{F}}}
\newcommand{\A}{{\mathcal{A}}}

\tikzfeynmanset{doublefermionarrow/.style={
                /tikz/double,
                /tikz/double distance = 1 pt,
                /tikz/decoration={name=none},
                /tikz/postaction={
                /tikzfeynman/with arrow=0.5,
                }
            }
        }

\tikzfeynmanset{doublefermion/.style={
                /tikz/double,
                /tikz/double distance = 1 pt,
                /tikz/decoration={name=none}
            }
        }

\tikzset{
	cut/.style={
		postaction={
			decorate,
			decoration={
				markings,
				mark=at position 1 with {
					\draw[thick] (0,-3.5pt) -- (0,3.5pt);
				}
			}
		}
	}, 
    dott/.style={
        postaction={
            decorate,
            decoration={
                markings,
                mark=at position 1 with {
                    \fill (0,0) circle (2pt);
                }
            }
        }
    },
	1piblob/.style={
		shape=circle,
		draw=black,
		fill=black!10,
		minimum size=0.7cm,
		inner sep=0pt,
		font=\footnotesize
	},
	shortphoton/.style={
		decorate,
		decoration={
			snake,
			amplitude=1.2pt,      
			segment length=3.5pt  
		}
	},
    otimes/.style={
        postaction={
            decorate,
            decoration={
                markings,
                mark=at position 1 with {
                    \draw[thick] (-2.5pt,-2.5pt) -- (2.5pt,2.5pt);
                    \draw[thick] (-2.5pt,2.5pt) -- (2.5pt,-2.5pt);
                }
            }
        }
    }
}

\tikzset{
    crossed square/.style={
        rectangle,
        draw,
        minimum width=8pt,
        minimum height=8pt,
        inner sep=0pt,
        path picture={
            \draw
            (path picture bounding box.north west) --
            (path picture bounding box.south east);
            \draw
            (path picture bounding box.south west) --
            (path picture bounding box.north east);
        }
    }
}

\tikzset{
    crossed triangle down/.style={
        regular polygon,
        regular polygon sides=3,
        shape border rotate=180,
        draw,
        fill=white,
        minimum size=3pt,
        inner sep=-2.1pt,
    }
}

\begin{document}

\title{Renormalized perturbation theory in an intense background electromagnetic field}
\author{Misha A. \surname{Lopez-Lopez}}
\email{mlopez28@pas.rochester.edu}
\affiliation{Department of Physics and Astronomy, University of Rochester, Rochester, New York 14627, USA}
\author{Giulio \surname{Audagnotto}}
\affiliation{Department of Physics and Astronomy, University of Rochester, Rochester, New York 14627, USA}
\affiliation{Dipartimento di Fisica, Universit\`a di Torino \& INFN, Sezione di Torino, \\
Via Pietro Giuria 1, I-10125 Turin, Italy}
\author{Antonino \surname{Di Piazza}}
\email{a.dipiazza@rochester.edu}
\affiliation{Department of Physics and Astronomy, University of Rochester, Rochester, New York 14627, USA}
\affiliation{Laboratory for Laser Energetics, University of Rochester, Rochester, New York 14623, USA}

\date{\today}

\begin{abstract}
Quantum electrodynamics in strong background electromagnetic fields or strong-field QED (SFQED) has been investigated in great detail at the tree level. The study of SFQED at higher loops has not been carried out in a correspondingly systematic way, with the notable exception of strong background atomic fields. In this work, we investigate the renormalization of SFQED by writing the standard unrenormalized SFQED Lagrangian density, which differs from the vacuum-QED Lagrangian density by the additional interaction term of the Dirac four-current density with the background four-vector potential, in terms of renormalized quantities and counterterms. Within this framework, we confirm the necessity of renormalizing the background field as an electric charge rather than as a photon field. We compare this approach with an alternative, less common one proposed in the literature, which features a different renormalized Lagrangian density, and we show that these two approaches are physically equivalent. The equivalence is demonstrated by showing that both Lagrangian densities can be derived from the same Lagrangian density, where the presence of the background field is described by the corresponding background four-current density. As a byproduct, it will become clear that the renormalizability of SFQED directly derives from the renormalizability of vacuum QED. By following the approach based on the standard unrenormalized SFQED Lagrangian density, we obtain that the renormalized SFQED Lagrangian density features a new counterterm as compared to the vacuum Lagrangian density. This counterterm is necessary to renormalize the electron self energy, which undergoes a new contribution as compared to vacuum QED, physically due to the electromagnetic field produced by the four-current density induced in the vacuum by the background electromagnetic field. The vacuum-induced electromagnetic field is generally a nonlinear function of the background electromagnetic field, but only the linear term in the background field diverges and needs to be renormalized. We study the corresponding counterterm in detail and we show how the renormalization of the vacuum-induced electromagnetic field is implemented explicitly at one loop and at two loops. The renormalization at all loops is shown to be more easily deduced at the level of the Lagrangian density. Finally, some subtleties concerning the renormalization procedure in the case of a free background field and, in particular, of a plane-wave field are also pointed out.
\end{abstract}

\maketitle

\section{Introduction}
%
QED in the presence of strong background electromagnetic fields has become a subject of great interest due to the developments on high-intensity lasers that can produce electromagnetic fields of unprecedented strengths. This will enable to systematically test QED in the so-called strong-field regime, where nonlinear and nonperturbative effects in the background field are expected to be dominant \cite{Di_Piazza_2012,Gonoskov_2022,Fedotov_2023}. In the strong-field QED (SFQED) regime processes occur in the presence of background electromagnetic fields of effective strengths of the order of the so-called Schwinger critical field $F_{cr} = m^2 /|e| = 1.3 \times 10^{16} \, \textrm{V/cm}=4.4\times 10^{13}\;\text{G}$ \cite{Schwinger_1951,Ritus_1985,Baier_b_1998,Landau_b_4_1982}, at which nonperturbative quantum phenomena become important (units with $\epsilon_0=c=\hbar=1$ are used throughout this paper such that the fine-structure constant is $\alpha=e^2/4\pi\approx 1/137$, and $m$ and $e<0$ are the (physical) electron mass and charge, respectively).

Current \cite{ELI,APOLLON_10P,CoReLS} and upcoming \cite{NSF_OPAL,Vulcan_20-20,SEL} laser facilities aim to produce beams with intensities in the range $10^{22}\text{-}10^{24}\, \textrm{W}/\textrm{cm}^2$. Notably, the record peak intensity of $1.1 \times 10^{23}\, \textrm{W}/\textrm{cm}^2$ has been already achieved \cite{CoReLS,Yoon_2021}. For a laser to reach the critical field limit, it would require a peak intensity of about $4.6\times 10^{29}\, \textrm{W}/\textrm{cm}^2$, which is still several orders of magnitude beyond current capabilities. However, it is possible nowadays to access the SFQED regime by colliding a high-intensity laser beam with, for example, high-energy (GeV or higher) electrons and/or positrons \cite{Di_Piazza_2012,Gonoskov_2022,Fedotov_2023} because, due to the Lorentz invariance of QED, physical observables depend on the value of the electromagnetic field strength in the rest frames of the charges. In fact, nonlinear Compton scattering, i.e., the photon emission by an electron or a positron in a strong laser field with sizable recoil, has been observed in the laboratory, by colliding a 45-GeV electron beam with a counter-propagating laser field of intensity of the order of $10^{18}\, \textrm{W}/\textrm{cm}^2$ \cite{Bula_1996}. In a related experimental campaign, nonlinear trident pair production, i.e., the decay into an electron-positron pair of a photon emitted by an electron colliding with a strong laser field, has also been observed  \cite{Burke:1997ew,Bamber:1999zt}. These experiments confirmed the predictions of SFQED but have been carried out in a regime where an electron/positron typically absorbs a few photons from the laser field during each process. More recent experiments have been performed in a regime where the emission of radiation by ultrarelativistic electrons occurs with the absorption of several photons from the laser field, leading to a highly-nonlinear dependence of the processes probabilities on the laser-field amplitude \cite{Cole_2018,Poder_2018,Mirzaie_2024,Los_2026}. 

SFQED processes can also be directly primed by photons like nonlinear Breit-Wheeler pair production, i.e., the decay of a photon into an electron-positron pair inside a strong laser field. Photons with energies of the order of GeV or higher are required for the decay in a laser field of intensity of the order of $10^{22}\text{-}10^{23}\, \textrm{W}/\textrm{cm}^2$ not to be exponentially suppressed \cite{Di_Piazza_2012,Gonoskov_2022,Fedotov_2023}. Correspondingly, detailed theoretical studies of the leading contributions to nonlinear Compton scattering and to nonlinear Breit-Wheeler pair production have been carried out (see the reviews \cite{Di_Piazza_2012,Gonoskov_2022,Fedotov_2023} and the references therein).

In order to test SFQED, it is essential to have a theoretical approach in which the interaction of electrons and positrons with the intense background field is described accurately. This requires solving the Dirac equation in the presence of the background field exactly and then quantize the electron-positron field by accounting for the background field exactly. This is not possible for an arbitrary electromagnetic field. A useful approximation, which has provided a significant insight in processes occurring in intense laser field is the plane-wave approximation, which is valid if the laser energy is not too tightly focused in space \cite{Mitter_1975,Ritus_1985,Ehlotzky_2009,Di_Piazza_2012,Gonoskov_2022,Fedotov_2023}. In this case the Dirac equation can be solved exactly and the corresponding electron states are called Volkov states \cite{Volkov_1935,Landau_b_4_1982}. Analogously, the analytical solution of the Dirac equation in the presence of a Coulomb field has played an important role in the study of SFQED in strong atomic fields \cite{Landau_b_4_1982,Shabaev_2024}.

Apart from the most recent experiments \cite{Cole_2018,Poder_2018,Mirzaie_2024,Los_2026}, which employed electron beams produced via laser wake-field acceleration to study SFQED in a strong laser field, two experimental campaigns aim at testing SFQED with high accuracy and, to this end, employing more controllable laser conditions and an electron beam produced by a conventional accelerator \cite{Chen_2022,Abramowicz_2024}. This together with more fundamental reasons motivates the study of radiative corrections in SFQED in the presence of strong laser fields and, in general, of the role of renormalization in the calculation of higher-order transition amplitudes in SFQED. We refer the reader to Refs. \cite{Braun_1972,Braun_1978,Brouder_2002} for previous studies on the renormalization procedure developed for QED in external fields (see also the review \cite{Shabaev_2024} in the case of background atomic fields). Below, we study the renormalization of SFQED including the case of background electromagnetic fields in which the vacuum is unstable \cite{Fradkin_b_1991}. As in the monograph \cite{Fradkin_b_1991}, we assume that it is possible to classify the states of the Dirac field with positive and negative energies in both the asymptotic past and future. This approach also includes classes of background fields for which the vacuum is unstable under electron-positron pair production. If the vacuum is stable in the background under consideration, the easier approach based on the so-called Furry picture can be employed \cite{Furry_1951,Landau_b_4_1982}, where a unique classification of the states of the Dirac field with positive and negative energies can be achieved for all times. We will consider as standard SFQED Lagrangian density the unrenormalized Lagrangian density of vacuum QED with the addition of the interaction term of the Dirac four-current density with the background four-vector potential.

As we will also explain below, the renormalization of SFQED is ultimately derived from that in vacuum QED although technically more complicated. One can show, for example, that by imposing quite natural renormalization conditions the values of the renormalization constants coincide with those in vacuum (or, more precisely, at least their divergent parts). This occurrence can be physically understood from the fact that renormalization cures the ultraviolet divergences of the theory, and at higher and higher energies one would expect that a background electromagnetic field does affect less and less the dynamics of the particles. Among others, a genuine novelty in the renormalization program of SFQED is represented by the so-called tadpole diagram \cite{Braun_1972,Braun_1978,Brouder_2002}, which is essentially the one-photon correlation function, which identically vanishes in vacuum due to Furry theorem or, equivalently, to charge-parity conservation \cite{Itzykson_b_1980}. The tadpole, as we will see, diverges and needs to be renormalized, but it does not require a new renormalization constant. It is important, however, that the background electromagnetic field is renormalized not as an electromagnetic field but as a charge. 

This method of renormalizing QED in an external field for a spin one-half particle was presented in Refs. \cite{Braun_1972,Braun_1978}, where after considering the renormalization of the tadpole contributions \cite{Braun_1972}, the relation that defines the renormalized background field in terms of the unrenormalized one was derived. The method had already been applied before to the case of a background Coulomb field (see below) and it is mentioned in the monographs \cite{Schweber_b_1989,Dyson_b_2011}. The whole renormalization procedure was then presented in Ref. \cite{Braun_1978}. In Ref. \cite{Brouder_2002}, the renormalization of QED in an external field was carefully studied with a different approach based on a path-integral formulation and on the Dyson-Schwinger equations. This topic has also been discussed in some textbooks \cite{Weinberg_b_1_2005,Collins_b_1984} using a different approach, which we will show below to be equivalent to that in Refs. \cite{Braun_1978,Brouder_2002}.

As we have already mentioned, for a consistent treatment of the renormalization of SFQED, one has to consider also diagrams containing tadpoles, which are also known especially in the community studying SFQED in highly-charged ions, as vacuum polarization. The reason for this nomenclature is that physically the external field ``polarizes'' the vacuum by generating a vacuum four-current density, which in turn produces a quantum correction to the classical background field, whose amplitude corresponds to the tadpole diagram. The result is that only the sum of the background field and the quantum correction effectively act on the charges. In the case of a Coulomb background field the one-loop quantum correction to the electromagnetic field has been computed in Ref. \cite{Uehling_1935} at the leading order with respect to the Coulomb field (Uehling potential) and in Refs. \cite{Wichmann_1956,Blomqvist:1972ddn,Gyulassy:1974ba,Soff:1988zz} at higher orders (Wichmann-Kroll corrections). At the two-loop order, the leading contribution in the Coulomb field \cite{Kallen_1955} and higher-order corrections \cite{Volkov_2025_a,Volkov_2025_b} have also been computed.

While it was well understood that in a Coulomb field tadpole corrections contribute, they were believed to vanish identically in a constant background field. This is because the vacuum four-current density in a constant background field vanishes as well \cite{Fradkin_b_1991,Brown_1964}. However, it was first shown in Ref. \cite{Gies_2017} that the two-loop one-particle reducible (1PR) contribution to the Euler-Heisenberg effective Lagrangian, featuring two vacuum four-current densities connected by a photon propagator, does not vanish in a constant field (recall that 1PR Feynman diagrams are connected diagrams that can be separated into two diagrams by cutting one internal line). This led to a systematic study of such tadpole-like contributions in constant fields, for the polarization operator \cite{Karbstein_2017}, for higher-order corrections to the Euler-Heisenberg Lagrangian \cite{Karbstein_2017,Karbstein:2019wmj}, for the electron propagator \cite{Edwards_2017,Ahmadiniaz_2017,Ahmadiniaz_2019}, for the case of charged particles with different flavors \cite{Karbstein:2021gdi}, and for the photon-graviton conversion \cite{Ahmadiniaz:2026yrw}. Finally, in the case of a plane-wave field, it was shown that all tadpole contributions vanish exactly after renormalization \cite{Ahmadiniaz_2019,Di_Piazza_2022_b}. 

Concerning loop corrections in SFQED, several advances have been made in Coulomb, constant-crossed, and plane-wave fields. The impressive accuracy in experiments with highly-charged ions (see e.g., Refs. \cite{Sturm_2014,Sailer_2022,Morgner-2023, Morgner_2025}) has motivated the nonperturbative calculation of the electron self energy and the vacuum polarization at one loop \cite{Mohr:1998grz,Yerokhin_2017,Yerokhin-2025-a} and two loops \cite{Yerokhin-2003,Yerokhin-2006,Yerokhin_2008,Yerokhin_2013,Yerokhin-2015,Indelicato-2019, Sikora_2020, Debierre_2021,Yerokhin_2024,Yerokhin-2025-b, Sikora_2025, Yerokhin_2025-c}. In particular, these corrections are essential in the comparison between theory and experiments for the Lamb shift and the bound-electron $g$-factor in hydrogenlike ions (see also Refs. \cite{Mohr:1998grz,Shabaev_2024} and references therein). 

Several contributions to the mass and polarization operators in constant-crossed field have been computed at one-loop \cite{Narozhny_1969,Ritus_1970}, two-loop \cite{Ritus_1972,Ritus_1972_b,Morozov_1975}, and three-loop order \cite{Narozhny_1979,Narozhny_1980}. The one-loop vertex correction was calculated in Refs. \cite{Morozov_1981,Morozov_1981_b} and the resummation of higher-order corrections to the polarization operator has been investigated in Ref. \cite{Mironov_2020} in relation to the so-called Ritus-Narozhny conjecture \cite{Fedotov_2017}. More recently the cutting rules to use loop diagrams to compute total transition probabilities in a strong constant-crossed field have also been formulated \cite{Selivanov_2026}.

For an arbitrary plane-wave field only the one-loop basic diagrams have been computed. These include the mass operator \cite{Baier_1976_a,Di_Piazza_2021_e}, the polarization operator \cite{Baier_1976_b,Becker_1975,Meuren_2013}, and the vertex correction \cite{Di_Piazza_2020_b}. The three-point photon amplitude exact in a plane wave has also been investigated for all photons being real \cite{Di_Piazza_2007} and for two photons being real and one photon corresponding to an atomic field \cite{Di_Piazza_2008_c}. The general properties of the $n$-point photon amplitude exact in a plane wave were studied in Ref. \cite{Di_Piazza_2013}.

The present paper is organized as follows. In Sec. \ref{sec-furry}, we introduce the SFQED Lagrangian density we will work with and discuss the properties of tadpole corrections. In Sec. \ref{sec-ren-sfqed}, we explain the renormalization procedure in SFQED, show how such a procedure differs from that in vacuum especially for the renormalization of the tadpole and the electron self energy, and compare two different methods of renormalization. In Sec. \ref{sec-ren-ct}, we discuss the role of the renormalization constants in the calculations of radiative corrections. After making a few considerations on the tadpole in Sec. \ref{Remarks_Tadpole} especially in the presence of free background electromagnetic fields, the conclusions are presented in Sec. \ref{sec-conclusion}. The appendix \ref{sec-tadpole-pw}, contains a re-evaluation of the one-loop vacuum four-current density in an arbitrary plane-wave field.

Throughout this work, we use the metric tensor $\eta^{\mu\nu} = \text{diag}(+1,-1,-1,-1)$. We define $\hat{v} = \gamma^\mu v_\mu$ for an arbitrary four vector $v^\mu$, with $\gamma^\mu$ being the Dirac matrices, and $(ab) = a^\mu b_\mu$ for two generic four-vectors $a^\mu$ and $b^\mu$. 
%
\section{SFQED and the tadpole Feynman diagram}\label{sec-furry}
%
In this section, we present an introduction to the Lagrangian density, the functional generators, and the propagators in SFQED and then we focus on the properties of the tadpole Feynman diagram.

\subsection{The SFQED Lagrangian density}
The unrenormalized Lagrangian density of QED (or vacuum QED) is 
\begin{equation}
\label{L_V}
\mathcal{L}_{V,0} = \bar{\psi}_0(i \hat{\partial} - m_0 ) \psi_0 - \frac{1}{4} F_{0,\mu \nu} F_0^{\mu \nu} - \frac{1}{2\rho_0}(\partial A_0)^2 - e_0 \bar{\psi}_0 \hat{A}_0\psi_0,
\end{equation}
where $\psi_0(x)$ is the Dirac field describing electrons and positrons, $F_0^{\mu \nu}(x) = \partial^\mu A_0^\nu(x) - \partial^\nu A_0^\mu(x)$ is the electromagnetic field tensor describing photons, with $A_0^\mu(x)$ being the corresponding four-vector potential, and $\rho_0$ is the gauge-fixing parameter \cite{Itzykson_b_1980}. The index 0 indicates that the fields are unrenormalized, whereas the constants $m_0$, $e_0$, and $\rho_0$ are bare quantities (the gauge parameter $\rho_0$ does not undergo radiative corrections as it is not a physical quantity, nevertheless it is convenient to use the same terminology). Note, however, that quantities without the subscript ``0'' may turn out not to be finite.

Below, we will consider the study of QED phenomena in the presence of a strong background field such that the effects of the latter must be taken into account exactly in the calculations. We consider the possibility that the vacuum is unstable in the presence of the background field. Following the formalism presented in Ref. \cite{Fradkin_b_1991}, we assume that in the asymptotic past and future the structure of the background field allows one for a consistent definition and classification of the electron states with positive and negative energies, which then allows for the construction of the Fock in- and out-states of electrons and positrons. In particular, the in-vacuum state and the out-vacuum state will be denoted below as $\ket{0,\text{in}}$ and $\ket{0,\text{out}}$, respectively, and the fact that the vacuum can become unstable corresponds to the probability $|\braket{0,\text{out}|0,\text{in}}|^2$ being smaller than unity \cite{Fradkin_b_1991}. The presence of the background field is taken into account exactly in the construction of the states $\ket{0,\text{in}}$ and $\ket{0,\text{out}}$, whereas the interaction between the Dirac field $\psi_0(x)$ and the photon field $A_0^{\mu}(x)$ is ignored. Below, we indicate as $\ket{\Omega,\text{in}}$ and $\ket{\Omega,\text{out}}$ the corresponding exact vacuum in-state and exact vacuum out-state, respectively, which also include the interaction between $\psi_0(x)$ and $A_0^{\mu}(x)$ to all orders.

In order to construct the Lagrangian density of SFQED in the presence of the background field $\mathcal{A}_0^{\mu}(x)$ produced by the four-current density $\mathcal{J}_0^{\mu}(x)$, we start from the vacuum-QED Lagrangian density $\mathcal{L}_{V,0}$ and add the interaction term $-(\mathcal{J}_0(x)A_0(x))$ between the four-current density $\mathcal{J}_0^{\mu}(x)$ and the electromagnetic field $A_0^{\mu}(x)$ (the free Lagrangian density of the four-current density should also be added but it is a non-dynamical term and it is ignored here). In the resulting Lagrangian density $\mathcal{L}_{B,0}=\mathcal{L}_{V,0} - (\mathcal{J}_0 A_0)$, the electromagnetic field $A_0^{\mu}(x)$ also describes the background electromagnetic field and, for the sake of clarity, we indicate it as the total electromagnetic field $A_{T,0}^{\mu}(x)$:
\begin{equation}
\label{Lb_0}
\mathcal{L}_{B,0} =  \bar{\psi}_0(i \hat{\partial} -  m_0 ) \psi_0 - \frac{1}{4} F_{T,0,\mu \nu}F_{T,0}^{\mu \nu}  - \frac{1}{2\rho_0}(\partial A_{T,0})^2
- e_0 \bar{\psi}_0\hat{A}_{T,0}\psi_0 - (\mathcal{J}_0 A_{T,0}),
\end{equation}
with $F_{T,0}^{\mu \nu}(x)=\partial^{\mu}A_{T,0}^{\nu}(x)-\partial^{\nu}A_{T,0}^{\mu}(x)$. At this point we make the substitution $A_{T,0}^\mu(x)=A_0^{\mu}(x) + \mathcal{A}_0^\mu(x)$ and use the assumption that the background four-current density $\mathcal{J}_0^{\mu}(x)$ and field $\mathcal{A}_0^{\mu}(x)$ are not dynamical quantities but rather given functions. The formula can be additionally simplified by assuming without loss of generality that the background field satisfies the Lorenz-gauge condition $(\partial \mathcal{A}_0(x))=0$. By ignoring all the non-dynamical terms in $\mathcal{L}_{B,0}$, we can rewrite it as
\begin{equation}
\label{L_B_0}
\mathcal{L}_{B,0} =  \bar{\psi}_0(i \hat{\partial}-e_0 \hat{\mathcal{A}}_0 -  m_0 ) \psi_0 - \frac{1}{4} F_{0,\mu \nu}F_0^{\mu \nu}- \frac{1}{2} F_{0,\mu \nu}\mathcal{F}_0^{\mu \nu}  - \frac{1}{2\rho_0}(\partial A_0)^2- e_0 \bar{\psi}_0\hat{A}_0\psi_0 
 - (\mathcal{J}_0 A_0),
\end{equation}
where $\mathcal{F}_0^{\mu\nu}(x)=\partial^{\mu}\mathcal{A}_0^{\nu}(x)-\partial^{\nu}\mathcal{A}_0^{\mu}(x)$. 

Finally, we assume that the background electromagnetic field $\mathcal{A}_0^{\mu}(x)$ and the background four-current density $\mathcal{J}_0^{\mu}(x)$ are related by Maxwell's equations $\partial_{\mu}\mathcal{F}_0^{\mu\nu}(x)=\mathcal{J}_0^{\nu}(x)$ (as we will discuss in Sec. \ref{sub-sec Collins and Weinberg}, this is not the only possible choice because what is physically important is that the renormalized background field satisfies the classical Maxwell's equations with the renormalized background four-current). This implies that the third and the last terms in Eq. (\ref{L_B_0}) form a total four-divergence and can also be ignored. The resulting Lagrangian density
\begin{equation}
\label{Lb}
\mathcal{L}_{B,0} =  \bar{\psi}_0(i \hat{\partial}-e_0 \hat{\mathcal{A}}_0 - m_0 ) \psi_0
- \frac{1}{4} F_{0,\mu \nu} F_0^{\mu \nu} - \frac{1}{2\rho_0}(\partial A_0)^2
- e_0 \bar{\psi}_0 \hat{A}_0 \psi_0
\end{equation}
is customarily referred to as the unrenormalized SFQED Lagrangian density. Below, the symbol $\mathcal{L}_{B,0}$ will refer to this expression of the unrenormalized Lagrangian density of SFQED, unless otherwise specified.

The equations of motion of the two dynamical fields $\psi_0(x)$ and $A_0^{\mu}(x)$ obtained from this Lagrangian density are
\begin{align}
\label{Eq_M_psi}
(i \hat{\partial}-e_0 \hat{\mathcal{A}}_0 - m_0)\psi_0  -  e_0 \hat{A}_0 \psi_0&= 0, \\
\label{Eq_M_A}
\left[\square  \eta^{\mu \nu} - \left(1-\frac{1}{\rho_0} \right) \partial^\mu \, \partial^\nu\right] A_{0,\nu} -  e_0\bar{\psi}_0\gamma^\mu \psi_0  &= 0,
\end{align} 
where $\square=\partial_{\mu}\partial^{\mu}$. 

The Lagrangian density in Eq. \eqref{Lb} is not the only form in which the Lagrangian density of SFQED can be found in the literature. For instance, in Ref. \cite{Brouder_2002} the Lagrangian density of SFQED is defined as the one in Eq. (\ref{Lb_0}). The resulting equations of motion 
\begin{align}
\label{Brouder_Eq_M_psi}
(i \hat{\partial} - m_0)\psi_0  -  e_0 \hat{A}_{T,0} \psi_0&= 0, \\
\label{Brouder_Eq_M_A}
\left[\square  \eta^{\mu \nu} - \left(1-\frac{1}{\rho_0} \right) \partial^\mu \, \partial^\nu\right]A_{T,0,\nu} -  e_0\bar{\psi}_0\gamma^\mu \psi_0  &= \mathcal{J}_0^{\mu}
\end{align} 
are indeed identical to Eqs. (\ref{Eq_M_psi}) and (\ref{Eq_M_A}) once the substitution $A_{T,0}^{\mu}(x)=A_0^{\mu}(x)+\mathcal{A}_0^{\mu}(x)$ is made, which confirms that the Lagrangian densities in Eqs. (\ref{Lb_0}) and (\ref{Lb}) are physically equivalent. Interestingly, in Ref. \cite{Fradkin_b_1991} the unitary operator was found, which connects the states of the theories with the two Lagrangian densities, with the field $\mathcal{A}_0^{\mu}(x)$ being expressed in terms of the four-current density $\mathcal{J}_0^{\mu}(x)$ via the retarded propagator of the wave operator. We will come back later to this point.

Now, the expression of the SFQED Lagrangian density in Eq. (\ref{Lb}) indicates that the presence of the background field implies the existence of a new vertex, with two fermion lines attached to a ``photon'' line corresponding to the background field. However, the background field is typically so intense that its effects must be taken into account exactly in the calculations. This is achieved by quantizing the Dirac field $\psi_0(x)$ in the presence of the background field itself \cite{Furry_1951,Landau_b_4_1982,Fradkin_b_1991}, whereas the electromagnetic field $A_0^{\mu}(x)$ is quantized as in vacuum via, e.g., the Gupta-Bleuler approach \cite{Itzykson_b_1980}, and only the interaction between the fields $\psi_0(x)$ and $A_0^{\mu}(x)$ (the last term in Eq. (\ref{Lb})) is treated perturbatively. In other words, one conceptually splits the SFQED Lagrangian density as $\mathcal{L}_{B,0}=\mathcal{L}_{B,f,0}+\mathcal{L}_{B,i,0}$, where
\begin{align}
\label{Lb-free}
\mathcal{L}_{B,f,0} &=  \bar{\psi}_0(i \hat{\partial}-e_0 \hat{\mathcal{A}}_0 - m_0 ) \psi_0 
- \frac{1}{4} F_{0,\mu \nu} F_0^{\mu \nu} - \frac{1}{2\rho_0}(\partial A_0)^2,\\
\mathcal{L}_{B,i,0} &=-e_0 \bar{\psi}_0 \hat{A}_0\psi_0,
\end{align}
are the ``free'' Lagrangian density (including the background electromagnetic field) and the interaction Lagrangian density, respectively. As we have mentioned in the Introduction, the quantization of the Dirac field relies on the possibility of solving analytically the Dirac equation in the presence of the background field $\mathcal{A}_0^{\mu}(x)$. This can be carried out only for highly symmetric fields, which include constant fields, plane-waves fields, and the Coulomb field \cite{Landau_b_4_1982,Fradkin_b_1991,Bagrov_b_2014}. 

\subsection{Functional generators and propagators in SFQED}
\label{sub-sec functionals}
In the following we will often employ functional methods, and it is thus useful to introduce here the basic mathematical tools of the functional approach to SFQED.

In vacuum QED, the generating functional of the unrenormalized Green's functions is defined as
\begin{equation}
\label{Z_0}
Z_0[0|J_0, \eta_0, \bar{\eta}_0] = \mathcal{N}_0\int \mathcal{D}\psi_0\mathcal{D}\bar{\psi}_0\mathcal{D}A_0\, e^{i\int d^4x\,(\mathcal{L}_{V,0} -(J_0 A_0) - \bar{\eta}_0 \psi_0 - \bar{\psi}_0 \eta_0)},
\end{equation}
where $\mathcal{N}_0$ is a normalization factor ensuring that the exact vacuum state $| \Omega \rangle$ in vacuum QED is stable and normalized to unity:
\begin{equation}
Z_0[0|0,0,0]= \langle \Omega | \Omega \rangle = 1
\end{equation}
and where $J^\mu_0(x)$, $\eta_0(x)$ and $\bar{\eta}_0(x)$ are the usual auxiliary sources. 
As it is well known, starting from this generator, the exact free fermion propagator
\begin{equation}
G_0(x-y|0)=-i
\langle \Omega|\mathcal{T}(\psi_0(x)\bar{\psi}_0(y))|\Omega\rangle
=-i\left.
\frac{\delta^2  \log(Z_0[0|J_0, \eta_0, \bar{\eta}_0)]}{\delta \bar{\eta}_0(x) \delta {\eta}_0(y)}
\right\vert_{J_0 =\eta_0=\bar{\eta}_0=0 }
\end{equation}
and the exact free photon propagator
\begin{equation}
\label{D}
D^{\mu\nu}_0(x-y|0)=i
\langle \Omega|\mathcal{T}(A^{\mu}_0(x)A^{\nu}_0(y))|\Omega\rangle
=-i\left.
\frac{\delta^2  \log(Z_0[0|J_0, \eta_0, \bar{\eta}_0)]}{\delta J_{0,\mu}(x) \delta J_{0,\nu}(y)}
\right\vert_{J_0 =\eta_0=\bar{\eta}_0=0 }
\end{equation}
can be constructed (here all operators are to be intended in the Heisenberg representation) \cite{Itzykson_b_1980}. Below, however, we will mostly use these propagators at the tree level in the interaction between the Dirac field and the electromagnetic (or radiation) field, which will be indicated by the upper index $(0)$. The resulting propagators $G^{(0)}_0(x-y|0)$ and $D^{(0),\mu\nu}_0(x-y|0)$ have the well-known expressions \cite{Itzykson_b_1980}
\begin{align}
\label{G_0}
G^{(0)}_0(x-y|0)&=\int\frac{d^4p}{(2\pi)^4}\frac{\hat{p}+m_0}{p^2-m_0^2+i0}e^{-i(p(x-y))},\\
\label{Free_Ph_Pr}
D^{(0),\mu\nu}_0(x-y|0)&=\int\frac{d^4k}{(2\pi)^4}\left[\frac{\eta^{\mu\nu}}{k^2+i0}+(\rho_0-1)\frac{k^{\mu}k^{\nu}}{(k^2+i0)^2}\right]e^{-i(k(x-y))}
\end{align}
and satisfy the equations
\begin{align}
(i\hat{\partial}-m_0)G^{(0)}_0(x-y|0)&=\delta^4(x-y),\\
\label{Eq_Free_Ph_Pr}
\left[\square \eta^{\mu \nu} - \left(1-\frac{1}{\rho_0} \right) \partial^\mu \, \partial^\nu\right]D^{(0)}_{0,\nu\lambda}(x-y|0)&=-\delta^{\mu}{}_{\lambda}\delta^4(x-y),
\end{align}
respectively. Also, they will be represented diagrammatically as
\begin{align}
\begin{tikzpicture}[baseline={([yshift=-0.5ex]current bounding box.center)}]
        \begin{feynman}
            \vertex (a1);
            \vertex [right=1.8cm of a1] (a2);
            \diagram*{
                (a2) -- [fermion] (a1),
            };
        \end{feynman}
    \end{tikzpicture}
    &= 
    i G^{(0)}_0(x-y|0), \\
    \begin{tikzpicture}[baseline={([yshift=-0.5ex]current bounding box.center)}]
        \begin{feynman}
            \vertex (a1);
            \vertex [right=1.8cm of a1] (a2);
            \diagram*{
                (a1) -- [photon] (a2),
            };
        \end{feynman}
    \end{tikzpicture}
    &= 
    -i D^{(0),\mu\nu}_0(x-y|0),
\end{align}
respectively.

The corresponding generating functional in SFQED is defined in analogy with Eq. (\ref{Z_0}) by replacing the vacuum QED Lagrangian density $\mathcal{L}_{V,0}$ with the SFQED Lagrangian density $\mathcal{L}_{B,0}$:
\begin{equation}
\label{Z_0_B}
Z_0[\mathcal{A}_0|J_0, \eta_0, \bar{\eta}_0] = \mathcal{N}_0\int \mathcal{D}\psi_0\mathcal{D}\bar{\psi}_0\mathcal{D}A_0\, e^{i\int d^4x\,(\mathcal{L}_{B,0} -(J_0 A_0) - \bar{\eta}_0 \psi_0 - \bar{\psi}_0 \eta_0)},
\end{equation}
which features the same normalization factor $\mathcal{N}_0$ as in Eq. (\ref{Z_0}). Indeed, according to the discussion in the Introduction, by setting all the sources equal to zero one obtains the transition amplitude between the exact in-vacuum state $\ket{\Omega, \text{in}}$ and the exact out-vacuum state $\ket{\Omega, \text{out}}$ in the background field, which in general is not equal to unity (or to a pure phase) \cite{Fradkin_b_1991}:
\begin{equation}\label{persistence-apt}
Z_0[\mathcal{A}_0| 0, 0, 0] = \langle \Omega, \text{out} | \Omega, \text{in} \rangle .
\end{equation}

The unrenormalized exact dressed electron propagator, i.e., the exact electron propagator in the presence of the background electromagnetic field $\mathcal{A}_0^{\mu}(x)$, is defined as \cite{Fradkin_b_1991}
\begin{equation}
\label{G_0_Exact}
G_0(x,y|\mathcal{A}_0)=-i
\frac{\langle \Omega,\text{out}|\mathcal{T}(\psi_0(x)\bar{\psi}_0(y))|\Omega,\text{in}\rangle}{\langle \Omega,\text{out}|\Omega,\text{in}\rangle}
=-i\left.
\frac{\delta^2  \log(Z_0[\mathcal{A}_0 |J_0, \eta_0, \bar{\eta}_0)]}{\delta \bar{\eta}_0(x) \delta {\eta}_0(y)}
\right\vert_{J_0 =\eta_0=\bar{\eta}_0=0 },
\end{equation}
where the equalities $\langle \Omega,\text{out}|\psi_0(x)|\Omega,\text{in}\rangle|_{J_0 =\eta_0=\bar{\eta}_0=0 }=0$ and $\langle \Omega,\text{out}|\bar{\psi}_0(x)|\Omega,\text{in}\rangle|_{J_0 =\eta_0=\bar{\eta}_0=0 }=0$ have been used, implying that only connected diagrams contribute to $G_0(x,y|\mathcal{A}_0)$ \cite{Brouder_2002}. This propagator will be represented diagrammatically by a thick, continuous line:
\begin{equation}
\label{G_0_0}
\begin{tikzpicture}[baseline={([yshift=-0.5ex]current bounding box.center)}]
        \begin{feynman}
            \vertex (a1);
            \vertex [right=1.8cm of a1] (a2);
            \diagram*{
                (a2) -- [fermion, ultra thick] (a1),
            };
        \end{feynman}
    \end{tikzpicture}
    = 
    i G_0(x,y|\mathcal{A}_0) .
\end{equation}
The corresponding tree-level, dressed fermion propagator $G^{(0)}_0(x,y|\mathcal{A}_0)$ will be instead indicated by a double line
\begin{equation}
    \begin{tikzpicture}[baseline={([yshift=-0.5ex]current bounding box.center)}]
        \begin{feynman}
            \vertex (a1);
            \vertex [right=1.8cm of a1] (a2);
            \diagram*{
                (a2) -- [ doublefermionarrow] (a1),
            };
        \end{feynman}
    \end{tikzpicture}
    =iG_0^{(0)}(x,y\,|\,\mathcal A_0),
    \end{equation}
and it satisfies the dressed Dirac equation
\begin{equation}\label{G_00}
[i \hat{\partial}_x-e_0 \hat{\mathcal{A}}_0(x) - m_0]G^{(0)}_0(x,y|\mathcal{A}_{0})=\delta^4(x-y).
\end{equation}

It is worth observing that the generating functional of SFQED $Z_0[\mathcal{A}_0|J_0, \eta_0, \bar{\eta}_0]$ can be related to the generating functional in vacuum QED $Z_0[0|J_0, \eta_0, \bar{\eta}_0]$. In fact, by shifting the integrated electromagnetic field $A_0^{\mu}(x)$ according to $A_0^{\mu}(x)\to A_0^{\mu}(x)+\mathcal{A}_0^{\mu}(x)$ in the functional $Z_0[0|J_0+\mathcal{J}_0,\eta_0,\bar{\eta}_0]$, we obtain
\begin{equation}
\label{W_0_vac_1}
\begin{split}
Z_0[0|J_0+\mathcal{J}_0,\eta_0,\bar{\eta}_0]&=\mathcal{N}_0\int\mathcal{D}\psi_0\mathcal{D}\bar{\psi}_0\mathcal{D}A_0\,e^{i\int d^4x\,\left[\mathcal{L}_{B,0}-\frac{1}{4}\mathcal{F}_{0,\mu\nu}\mathcal{F}_0^{\mu\nu}-((J_0+\mathcal{J}_0)\mathcal{A}_0)-J_{0,\mu}A_0^{\mu}-\bar{\psi}_0\eta_0-\bar{\eta}_0\psi_0\right]}\\
&=Z_0[\mathcal{A}_0|J_0, \eta_0, \bar{\eta}_0]e^{-i\int d^4x\,\left[\frac{1}{4}\mathcal{F}_{0,\mu\nu}\mathcal{F}_0^{\mu\nu}+((J_0+\mathcal{J}_0)\mathcal{A}_0)\right]}\\
&=Z_0[\mathcal{A}_0|J_0, \eta_0, \bar{\eta}_0]e^{-i\int d^4x\,\left(-\frac{1}{4}\mathcal{F}_{0,\mu\nu}\mathcal{F}_0^{\mu\nu}+(J_0\mathcal{A}_0)\right)},
\end{split}
\end{equation}
where we employed integration by parts and we used Maxwell's equations for the background field and its generating four-current density. Notably, in order to obtain the generating functional of SFQED it is not sufficient to add the background four-current density to the generating functional of vacuum QED but an additional exponential functional arises. This has interesting consequences for the renormalization, which will be pointed out below.

\subsection{The tadpole and the total vacuum electromagnetic field}
\label{Tad_Tot}
\begin{figure}
\includegraphics[width=5cm]{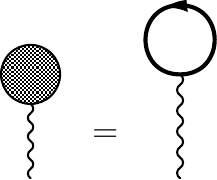}
\caption{Exact tadpole Feynman diagram representing the one-point correlation function of the electromagnetic field. The blob coincides with the exact dressed electron propagator in Eq. (\ref{G_0_Exact}) (see also Eq. (\ref{G_0_0})). The wavy line indicates a free, tree-level photon propagator.}
\label{Tadpole_Exact}
\end{figure}

An important difference between vacuum QED and SFQED is that the background electromagnetic field can induce a vacuum four-current density, which in turn generates an additional background field resulting in a total ``vacuum'' electromagnetic field which effectively interacts with the charges \cite{Braun_1972,Brouder_2002}. This purely quantum effect is known as vacuum polarization and is described diagrammatically by the so-called ``tadpole'' diagram in Fig. \ref{Tadpole_Exact}. More precisely, the tadpole diagram in Fig. \ref{Tadpole_Exact} without the photon line corresponds to the vacuum four-current density induced by the background electromagnetic field, whereas the diagram with the photon line corresponds to the electromagnetic field due to the vacuum polarization. Note that the photon line is treated as an internal line, which corresponds to a free photon propagator in the corresponding Feynman amplitude.

After the quantization of the Dirac field, the canonical operator $j_{c,0}^{\mu}(x)=e_0\bar{\psi}_0(x)\gamma^{\mu}\psi_0(x)$ of the conserved Dirac four-current density has the disadvantage to change sign only up to an infinite constant under charge conjugation \cite{Bjorken_b_1965}. This inconvenience is solved by re-defining the four-current density as $j_0^{\mu}(x)=(e_0/2)[\bar{\psi}_0(x),\gamma^{\mu}\psi_0(x)]$, with the square-brackets indicating the commutator between the Dirac field and the Dirac conjugated field only \cite{Bjorken_b_1965}. In vacuum QED, a consequence of this replacement is that in applying Wick's theorem to compute the $S$-matrix, the contraction of the fields $\psi_0(x)$ and $\bar{\psi}_0(x)$ belonging to the same Hamiltonian density identically vanishes. By using the same definition of $j_0^{\mu}(x)$ in SFQED, however, the contraction of the fields $\psi_0(x)$ and $\bar{\psi}_0(x)$ belonging to the same Hamiltonian density does not vanish and corresponds to the tadpole diagram computed via the symmetric time limit of the Feynman propagator according to Eq. (\ref{T_0}) below. For this reason, for the purpose of computing transition amplitudes also at higher orders in $\alpha_0$, one can equivalently use the canonical expression $j_{c,0}^{\mu}(x)$ of the four-current density, provided that in applying Wick's theorem the contractions of the Dirac field and its Dirac conjugated with the same coordinates are included and computed accordingly.

The tadpole amplitude $T_0^{\mu}(x)$, corresponding to the diagram in Fig. \ref{Tadpole_Exact} without the photon line, is given by
\begin{equation}
\label{T_0}
T_0^{\mu}(x)=-\text{Tr}((-ie_0\gamma^{\mu})iG_0(x,x|\mathcal{A}_0))=-e_0\text{Tr}(\gamma^{\mu}G_0(x,x|\mathcal{A}_0)),
\end{equation}
where the limit $y\to x$ has to be taken symmetrically for $y^0>x^0$ and for $y^0<x^0$,  see Ref. \cite{Schwinger_1951} and the appendix. The unrenormalized induced vacuum four-current density $\mathcal{J}_{v,0}^{\mu}(x)$ is defined as 
\begin{equation}
\label{J_v_0}
\mathcal{J}_{v,0}^{\mu}(x)=iT_0^{\mu}(x)=-ie_0\text{Tr}(\gamma^{\mu}G_0(x,x|\mathcal{A}_0))
\end{equation}
and \cite{Schwinger_1951,Fradkin_b_1991}
\begin{equation}
\label{J_v_0_av}
\mathcal{J}_{v,0}^{\mu}(x)=\frac{\langle \Omega,\text{out}|j^{\mu}_0(x)|\Omega,\text{in}\rangle}{\langle \Omega,\text{out}|\Omega,\text{in}\rangle}.
\end{equation}

By adding the photon line as an internal line, the tadpole diagram coincides with the one-point correlation function of the electromagnetic field, which is defined as (see Eq. (\ref{Z_0_B}) and Fig. \ref{Tadpole_Exact})
\begin{equation}
\label{A_v_0}
\frac{\langle \Omega,\text{out}| A^{\mu}_0(x)|\Omega,\text{in}\rangle}{\langle \Omega,\text{out}|\Omega,\text{in}\rangle}=i\left.
\frac{\delta  \log(Z_0[\mathcal{A}_0 |J_0, \eta_0, \bar{\eta}_0)]}{\delta J_{0,\mu}(x)}
\right\vert_{J_0 =\eta_0=\bar{\eta}_0=0 }
\end{equation}
and which can analogously be called the induced vacuum electromagnetic field:
\begin{equation}
\label{A_v_0_av}
\mathcal{A}_{v,0}^{\mu}(x)=\frac{\langle \Omega,\text{out}| A^{\mu}_0(x)|\Omega,\text{in}\rangle}{\langle \Omega,\text{out}|\Omega,\text{in}\rangle}.
\end{equation}
To show this equality, we use the identity
\begin{equation}
\begin{split}
0&=\int \mathcal{D}\psi_0\mathcal{D}\bar{\psi}_0\mathcal{D}A_0\, \frac{\delta}{\delta A_{0,\mu}(x)}e^{i\int d^4x\,(\mathcal{L}_{B,0} -(J_0A_0) - \bar{\eta}_0 \psi_0 - \bar{\psi}_0 \eta_0)}\\
&=i\int \mathcal{D}\psi_0\mathcal{D}\bar{\psi}_0\mathcal{D}A_0\, \left\{\left[\square \eta^{\mu \nu} - \left(1-\frac{1}{\rho_0} \right) \partial^\mu \, \partial^\nu\right]A_{0,\nu}(x)-e_0\bar{\psi}_0(x)\gamma^{\mu}\psi_0(x)-J^\mu_0(x)\right\}\\
&\quad\times e^{i\int d^4x\,(\mathcal{L}_{B,0} -(J_0A_0) - \bar{\eta}_0 \psi_0 - \bar{\psi}_0 \eta_0)}.
\end{split}
\end{equation}
By setting the sources equal to zero, we obtain
\begin{equation}
\left[\square \eta^{\mu \nu} - \left(1-\frac{1}{\rho_0} \right) \partial^\mu \, \partial^\nu\right]\mathcal{A}_{v,0,\nu}(x)=\mathcal{J}_{v,0}^{\mu}(x),
\end{equation}
where we have used the fact that, since the vacuum in-state $\ket{\Omega,\text{in}}$ and the vacuum out-state $\ket{\Omega,\text{out}}$ are physical states, according to the Gupta-Bleuler quantization procedure, the one-point correlation function of the electromagnetic field satisfies the Lorenz-gauge condition \cite{Mandl_b_2010}. By finally inverting the operator on the left-hand side according to the Feynman prescription, we have (see Eq. (\ref{Eq_Free_Ph_Pr}))
\begin{equation}
\label{A_v}
\mathcal{A}_{v,0}^{\mu}(x)=-\int d^4yD_0^{(0),\mu\nu}(x-y|0)\mathcal{J}_{v,0,\nu}(y)=ie_0\int d^4yD_0^{(0),\mu\nu}(x-y|0)\text{Tr}(\gamma_{\nu}G_0(y,y|\mathcal{A}_0)),
\end{equation}
which is exactly the amplitude corresponding to the Feynman diagram in Fig. \ref{Tadpole_Exact} including the photon line as an internal line.

Now, attaching the tadpole with the internal photon line to a fermion line in all possible ways is carried out in the same way as for the external background field such that one can consider together each background field line with each tadpole. In other words, one is led to introduce what we call the ``total vacuum electromagnetic field'', given by
\begin{equation}
\label{A_phys}
\begin{split}
\mathcal{A}^{\mu}_{T,0}(x)&=\mathcal{A}_0^{\mu}(x)+\mathcal{A}_{v,0}^{\mu}(x)=\mathcal{A}_0^{\mu}(x)-\int d^4yD_0^{(0),\mu\nu}(x-y|0)\mathcal{J}_{v,0,\nu}(y)\\
&=\mathcal{A}_0^{\mu}(x)+ie_0\int d^4yD_0^{(0),\mu\nu}(x-y|0)\text{Tr}(\gamma_{\nu}G_0(y,y|\mathcal{A}_0)),
\end{split}
\end{equation}
which is nothing but the one-point correlation function of the total electromagnetic field operator $A^{\mu}_{T,0}(x)=\mathcal{A}^{\mu}_0(x)+A^{\mu}_0(x)$, i.e.,
\begin{equation}\label{A_phys-2}
\mathcal{A}^{\mu}_{T,0}(x)=\frac{\langle \Omega,\text{out}| A^{\mu}_{T,0}(x)| \Omega,\text{in}\rangle}{\langle \Omega,\text{out}|\Omega,\text{in}\rangle}=\mathcal{A}^{\mu}_0(x)+\frac{\langle \Omega,\text{out}| A^{\mu}_0(x)| \Omega,\text{in}\rangle}{\langle \Omega,\text{out}|\Omega,\text{in}\rangle}.
\end{equation}
Note that the total vacuum electromagnetic field also satisfies the Lorenz-gauge condition $(\partial\mathcal{A}_{T,0}(x))=0$.
At least formally, the physical interpretation of the total vacuum electromagnetic field is that the (classical) background field $\mathcal{A}_0^{\mu}(x)$ ``polarizes'' the vacuum generating the (quantum) induced vacuum four-current density $\mathcal{J}_{v,0}^{\mu}(x)$, which in turn produces the (quantum) induced vacuum field $-\int d^4yD_0^{(0),\mu\nu}(x-y|0)\mathcal{J}_{v,0,\nu}(y)$. As a result, an electric charge can physically interact only with the sum of these two fields, i.e., with the total vacuum electromagnetic field $\mathcal{A}^{\mu}_{T,0}(x)$. The reason why we call this a formal physical interpretation is that, rigorously speaking, both the induced vacuum four-current density $\mathcal{J}_{v,0}^{\mu}(x)$ and the induced vacuum electromagnetic field $\mathcal{A}_{v,0}^{\mu}(x)$ (and then also the total vacuum electromagnetic field $\mathcal{A}_{T,0}^{\mu}(x)$) are not real quantities in a background field for which the vacuum state is unstable (see Eqs. (\ref{J_v_0_av}) and (\ref{A_v_0_av})). In fact, these quantities are actually correlators or Green's functions and, in order to be used in relation to physical transition amplitudes, all correlators have to be defined between the vacuum out-state and in-state starting from the generator $Z_0[\mathcal{A}_0|J_0, \eta_0, \bar{\eta}_0]$ in Eq. (\ref{Z_0_B}). Needless to say, one can also introduce average quantities like the total average electromagnetic field $\langle \Omega,\text{in}| A^{\mu}_{T,0}(x)| \Omega,\text{in}\rangle$ \cite{Fradkin_b_1991}, which is real, but these are not the quantities appearing, for example, in the Schwinger-Dyson equations derived from the generator $Z_0[\mathcal{A}_0|J_0, \eta_0, \bar{\eta}_0]$. It should also be stressed that these remarks on the difference between the introduced correlators and the corresponding real average values have far less impact experimentally as the vacuum instability is safely negligible for available electromagnetic fields. Also from a purely theoretical point of view, for widely-used background fields like a Coulomb field with charge number less than $137$, constant magnetic fields, and plane waves, the vacuum is rigorously stable \cite{Fradkin_b_1991} and the above interpretation holds rigorously.

Keeping in mind the previous observation, the interpretation of the total vacuum electromagnetic field $\mathcal{A}^{\mu}_{T,0}(x)$ as the electromagnetic field effectively interacting with the charges can be clarified by considering the Schwinger-Dyson equation of the unrenormalized exact dressed electron propagator $G_0(x,y|\mathcal{A}_0)$. This equation is most easily derived starting from the identity (see Eq. (\ref{Z_0_B}))
\begin{equation}
\label{Id_delta_psi_bar}
\begin{split}
0&=\int \mathcal{D}\psi_0\mathcal{D}\bar{\psi}_0\mathcal{D}A_0\, \frac{\delta}{\delta \bar{\psi}_0(x)}e^{i\int d^4x\,(\mathcal{L}_{B,0} -(J_0A_0) - \bar{\eta}_0 \psi_0 - \bar{\psi}_0 \eta_0)}\\
&=i\int \mathcal{D}\psi_0\mathcal{D}\bar{\psi}_0\mathcal{D}A_0\, \left\{[i\hat{\partial}_x-m_0-e_0\hat{A}_{T,0}(x)]\psi_0(x)-\eta_0(x)\right\}\\
&\quad\times e^{i\int d^4x\,(\mathcal{L}_{B,0} -(J_0A_0) - \bar{\eta}_0 \psi_0 - \bar{\psi}_0 \eta_0)}\\
&=i\left[i\hat{\partial}_x-e_0\hat{\mathcal{A}}_0(x)-m_0-ie_0\gamma^{\mu}\frac{\delta}{\delta J_0^{\mu}(x)}\right]\int \mathcal{D}\psi_0\mathcal{D}\bar{\psi}_0\mathcal{D}A_0\,\psi_0(x) e^{i\int d^4x\,(\mathcal{L}_{B,0} -(J_0A_0) - \bar{\eta}_0 \psi_0 - \bar{\psi}_0 \eta_0)}\\
&\quad-i\eta_0(x)Z_0[\mathcal{A}_0|J_0, \eta_0, \bar{\eta}_0]\\
&=-\left[i\hat{\partial}_x-e_0\hat{\mathcal{A}}_0(x)-m_0-ie_0\gamma^{\mu}\frac{\delta}{\delta J_0^{\mu}(x)}\right]Z_0[\mathcal{A}_0|J_0, \eta_0, \bar{\eta}_0]\frac{\delta\log(Z_0[\mathcal{A}_0|J_0, \eta_0, \bar{\eta}_0])}{\delta\bar{\eta}_0(x)}\\
&\quad-i\eta_0(x)Z_0[\mathcal{A}_0|J_0, \eta_0, \bar{\eta}_0].
\end{split}
\end{equation}
By taking the functional derivative $\delta/\delta\eta_0(y)$ of this equation and by setting all the sources to zero afterwards, one obtains (see also Eq. (\ref{G_0_Exact}), the equalities below it, and Eqs. (\ref{A_v_0}), (\ref{A_v_0_av}), and (\ref{A_phys-2})):
\begin{equation}
\label{Eq_Prop}
[i\hat{\partial}_x-e_0\hat{\mathcal{A}}_{T,0}(x)-m_0]G_0(x,y|\mathcal{A}_0)=\delta^4(x-y)-e_0\gamma^{\mu}\left.\frac{\delta^3 \log(Z_0[\mathcal{A}_0|J_0, \eta_0, \bar{\eta}_0])}{\delta J_0^\mu(x)\delta\eta_0(y)\delta\bar{\eta}_0(x)}\right\vert_{J_0=\eta_0=\bar{\eta}_0=0}.
\end{equation}
Now, we notice that the propagator $G_0(x,y|\mathcal{A}_0)$ is exact both with respect to the external field and with respect to the interaction between the Dirac field and the quantum electromagnetic field. In order to resemble the situation in vacuum QED, for a reason that will be clear below, it is convenient to introduce an alternative form of the exact dressed propagator, which we indicate as $G_{T,0}(x,y|\mathcal{A}_{T,0})$ and which is constructed starting from the corresponding tree-level propagator $G^{(0)}_{T,0}(x,y|\mathcal{A}_{T,0})$, namely the Green's function of the Dirac equation in the presence of the total vacuum electromagnetic field $\mathcal{A}^{\mu}_{T,0}(x)$ (see also Eqs. (\ref{G_00}) and (\ref{Eq_Prop})), i.e.,
\begin{equation}\label{G_00T}
[i \hat{\partial}_x-e_0 \hat{\mathcal{A}}_{T,0}(x) - m_0]G^{(0)}_{T,0}(x,y|\mathcal{A}_{T,0})=\delta^4(x-y).
\end{equation}
In other words, it is $G_0(x,y|\mathcal{A}_0)=G_{T,0}(x,y|\mathcal{A}_{T,0})$, but the higher-order corrections to the left-hand side and to the right-hand side of this inline equation are built up using electron lines which include exactly the interaction with the external field $\mathcal{A}^{\mu}_0(x)$ and with the total vacuum electromagnetic field $\mathcal{A}^{\mu}_{T,0}(x)$, respectively. This clearly implies that at the tree-level it is $G^{(0)}_0(x,y|\mathcal{A}_0)=G^{(0)}_{T,0}(x,y|\mathcal{A}_0)$. The advantage of using the exact propagator $G_{T,0}(x,y|\mathcal{A}_{T,0})$ is that its diagrammatic structure does not include explicitly the tadpole contributions, which are absorbed in the definition of $\mathcal{A}_{T,0}^{\mu}(x)$, and it is therefore formally identical to the exact electron propagator in vacuum (with the electron lines corresponding to propagators $G^{(0)}_{T,0}(x,y|\mathcal{A}_{T,0})$). Therefore, by also using Eq. (\ref{G_0_Exact}), we can conveniently rewrite Eq. (\ref{Eq_Prop}) as
\begin{equation}
[i\hat{\partial}_x-e_0\hat{\mathcal{A}}_{T,0}(x)-m_0]G_{T,0}(x,y|\mathcal{A}_{T,0})=\delta^4(x-y)-ie_0\gamma_{\mu}\int d^4z\frac{\delta G_{T,0}(x,y|\mathcal{A}_{T,0})}{\delta \mathcal{A}_{T,0}^{\nu}(z)}D^{\mu\nu}_{T,0}(x,z|\mathcal{A}_{T,0}),
\end{equation}
where we have introduced the unrenormalized exact dressed photon propagator $D_{T,0}^{\mu\nu}(x,y|\mathcal{A}_{T,0})$  in terms of the total vacuum electromagnetic field (see also Eqs. (\ref{A_v_0}) and (\ref{A_phys-2})) \cite{Brouder_2002}, which coincides with the exact photon propagator in the background field $\mathcal{A}^{\mu}_0(x)$:
\begin{equation}
D_{0}^{\mu\nu}(x,y|\mathcal{A}_{0})=-i\left.\frac{\delta^2 \log(Z_0[\mathcal{A}_0|J_0, \eta_0, \bar{\eta}_0])}{\delta J_{0,\nu}(y)\delta J_{0,\mu}(x)}\right\vert_{J_0=\eta_0=\bar{\eta}_0=0}.
\end{equation}
Note, in particular, the  symmetry of $D_{0}^{\mu\nu}(x,y|\mathcal{A}_{0})$ under the combined exchange of the Lorentz indices and the spacetime variables.

At this point, we use the general identity
\begin{equation}
\frac{\delta G_{T,0}(x,y|\mathcal{A}_{T,0})}{\delta \mathcal{A}_{T,0}^{\nu}(z)}=-\int d^4r\, d^4s\,G_{T,0}(x,r|\mathcal{A}_{T,0})\frac{\delta G^{-1}_{T,0}(r,s|\mathcal{A}_{T,0})}{\delta \mathcal{A}_{T,0}^{\nu}(z)}G_{T,0}(s,y|\mathcal{A}_{T,0})
\end{equation}
and the relation
\begin{equation}
\frac{\delta G^{-1}_{T,0}(r,s|\mathcal{A}_{T,0})}{\delta \mathcal{A}_{T,0}^{\nu}(z)}=-e_0\Lambda_{T,0,\nu}(z,r,s|\mathcal{A}_{T,0})
\end{equation}
between the derivative of the inverse of the dressed exact propagator and the exact vertex \cite{Itzykson_b_1980} to obtain the final equation
\begin{equation}
\label{Eq_Prop_f}
[i\hat{\partial}_x-e_0\hat{\mathcal{A}}_{T,0}(x)-m_0]G_{T,0}(x,y|\mathcal{A}_{T,0})=\delta^4(x-y)+\int d^4s\,\Sigma_{T,0}(x,s|\mathcal{A}_{T,0})G_{T,0}(s,y|\mathcal{A}_{T,0}).
\end{equation}
In this equation the unrenormalized dressed mass operator $\Sigma_{T,0}(x,y|\mathcal{A}_{T,0})$, defined according to the relation \cite{Itzykson_b_1980}
\begin{equation}
\begin{split}
-i\Sigma_{T,0}(x,y|\mathcal{A}_{T,0})&=-ie_0\gamma_{\mu}\int d^4z\, d^4r(-i)D^{\mu\nu}_{T,0}(x,z|\mathcal{A}_{T,0})iG_{T,0}(x,r|\mathcal{A}_{T,0})\\
&\quad\times(-ie_0)\Lambda_{T,0,\nu}(z,r,y|\mathcal{A}_{T,0})\\
&=-e^2_0\gamma_{\mu}\int d^4z\,d^4rD^{\mu\nu}_{T,0}(x,z|\mathcal{A}_{T,0})G_{T,0}(x,r|\mathcal{A}_{T,0})\Lambda_{T,0,\nu}(z,r,y|\mathcal{A}_{T,0}),
\end{split}
\end{equation}
has been introduced. We stress again that the  propagators, the vertex, and the mass correction are expressed in terms of the total vacuum electromagnetic field such that the higher-order corrections have the same diagrammatic structure as the corresponding quantities in vacuum. Equation (\ref{Eq_Prop_f}) shows that the effective electromagnetic field interacting with the charges as a classical background field is the total electromagnetic field $\mathcal{A}_{T,0}^{\mu}(x)$, whereas the additional term on the right-hand side features the mass operator. Alternatively, one can split the total vacuum electromagnetic field into the background field plus the quantum correction due to the tadpole, bring the latter correction to the right-hand side of Eq. (\ref{Eq_Prop_f}), and conclude that the exact dressed electron propagator undergoes corrections not only due to the mass operator, as in vacuum, but also to the tadpole, i.e., to the induced quantum vacuum field. Within this interpretation the tadpole is seen as an additional contribution to the electron self energy.

We conclude this paragraph by observing that one might wonder why the induced electromagnetic field in Eq. (\ref{A_phys}) is defined by using the Feynman propagator $D^{(0)}_{0,\nu\lambda}(x-y|0)$ rather than, as one would expect physically, the retarded propagator. The same question would arise concerning the background electromagnetic field and the background four-current density. In fact, we have never needed to express the background field in terms of the background four-current density as we have only exploited the fact that they are connected via Maxwell's equations. However, we would like to discuss this point because if one would start from the SFQED Lagrangian density in Eq. (\ref{Lb_0}) (see also Ref. \cite{Brouder_2002}), one would obtain the generating functional
\begin{equation}
\label{Z_0_J_0}
Z_0[0|J_0+\mathcal{J}_0,\eta_0,\bar{\eta}_0] = \mathcal{N}_0\int \mathcal{D}\psi_0\mathcal{D}\bar{\psi}_0\mathcal{D}A_{T,0}\, e^{i\int d^4x\,[\mathcal{L}_{V,0} -((J_0+\mathcal{J}_0)A_{T,0}) - \bar{\eta}_0 \psi_0 - \bar{\psi}_0 \eta_0]},
\end{equation}
where the vacuum-QED Lagrangian density $\mathcal{L}_{V,0}$ is given by Eq. (\ref{L_V}) with $A_0^{\mu}(x)\to A_{T,0}^{\mu}(x)$ and $F_0^{\mu\nu}(x)\to F_{T,0}^{\mu\nu}(x)$. By computing the total vacuum electromagnetic field $\A^\mu_{T,0}(x)$, one obtains that at the tree-level in the interaction between the Dirac field and the electromagnetic field
\begin{equation}
\A^{(0),\mu}_{T,0}(x)=-\int d^4y\,D_0^{(0),\mu\nu}(x-y|0)\mathcal{J}_{0,\nu}(y),
\end{equation}
because the integrand of the path integral in Eq. \eqref{Z_0_J_0} becomes Gaussian and the convergence of the path integral imposes the Feynman prescription for the propagator \cite{Itzykson_b_1980}.

These results are a consequence of expanding the $S$-matrix by using Wick's theorem and then applying Feynman rules to compute the correlators like $\mathcal{A}_{T,0}^{\mu}(x)$ (see Eq. (\ref{A_phys})), which imply the use of the Feynman propagator to obtain ultimately consistent and causal results. Indeed, the original expression of the $S$-matrix according to the Dyson formula is manifestly causal and it is only once it is expressed in terms of the time-ordered product of operators, that it looses its manifest causality \cite{Itzykson_b_1980}. As we have already discussed below Eq. (\ref{Brouder_Eq_M_A}), the equivalence of the SFQED Lagrangian densities used here and in Ref. \cite{Brouder_2002} has been explicitly proven when the background field and the four-current density are related via the retarded propagator \cite{Fradkin_b_1991}.

\section{Renormalization of SFQED}\label{sec-ren-sfqed}
As we have already noticed, all the quantities introduced in the previous sections are either unrenormalized fields (the quantum ones $\psi_0(x)$, $A_0^{\mu}(x)$, and the background one $\mathcal{A}_0^{\mu}(x)$) or bare constants ($m_0$, $e_0$, and $\rho_0$). Since our aim is to provide a recipe on how to compute SFQED correlation functions beyond the leading order in $\alpha_0$, we have to face the onset of ultraviolet divergences, which require renormalization. In this paper, we will not discuss neither the so-called Lehmann-Symanzik-Zimmermann reduction formula to construct transition amplitudes from the correlation functions nor infrared divergences, which are non-trivial problems on their own if treated to all orders in $\alpha_0$ within SFQED.

In order to establish how to define the renormalized fields and constants in terms of the corresponding unrenormalized quantities, we recall that, since ultraviolet divergences arise at higher and higher energies where the background field is expected not to significantly influence the particles' dynamics, we envisage a close relation between the renormalization in vacuum QED and in SFQED. As it was already noticed in Ref. \cite{Braun_1978}, this expectation can be based on solid mathematical grounds by referring to Eq. (\ref{W_0_vac_1}), which we rewrite in terms of the generators $W_0[0|J_0+\mathcal{J}_0,\eta_0,\bar{\eta}_0]=-i\log(Z_0[0|J_0+\mathcal{J}_0,\eta_0,\bar{\eta}_0])$ and $W_0[\mathcal{A}_0|J_0,\eta_0,\bar{\eta}_0]=-i\log(Z_0[\mathcal{A}_0|J_0,\eta_0,\bar{\eta}_0])$ of the unrenormalized connected Feynman diagrams as \cite{Itzykson_b_1980}
\begin{equation}
\label{W_0_J_W_0}
W_0[0|J_0+\mathcal{J}_0,\eta_0,\bar{\eta}_0]=W_0[\mathcal{A}_0|J_0,\eta_0,\bar{\eta}_0]+\int d^4 x\,\left[\frac{1}{4}\mathcal{F}_{0,\mu\nu}(x)\mathcal{F}_0^{\mu\nu}(x)-(J_0(x)\mathcal{A}_0(x))\right].
\end{equation}
Now, the left-hand side of this equation features the unrenormalized generator of vacuum QED, with an additional background four-current $\mathcal{J}_0^{\mu}(x)$ summed to the auxiliary source $J_0^{\mu}(x)$ of the electromagnetic field. Thus, defining the renormalized background four-current density as the renormalized auxiliary source of the electromagnetic field, i.e., setting $\mathcal{J}^{\mu}_0(x) = \mathcal{J}^{\mu}(x)/\sqrt{Z_3} $ and $J^{\mu}_0(x) = J^{\mu}(x)/\sqrt{Z_3}$, together with $\eta_0(x)=\eta(x)/\sqrt{Z_2}$ and $\bar{\eta}_0(x)=\bar{\eta}(x)/\sqrt{Z_2}$, guarantees that the resulting renormalized generator $W[0|J+ \mathcal{J},\eta,\bar{\eta}]=W_0[0|J/\sqrt{Z_3}+ \mathcal{J}/\sqrt{Z_3},\eta/\sqrt{Z_2},\bar{\eta}/\sqrt{Z_2}]$ is finite and generates finite correlation functions. Using the same relations between renormalized and unrenormalized quantities as in vacuum QED then renders the right-hand side of Eq. (\ref{W_0_J_W_0}) also finite. This determines how the Dirac and the electromagnetic fields as well as the electron mass and charge renormalize (apart from the unphysical gauge-fixing parameter). Thus, we still have to work out the renormalization of the background field, which can be carried out via the following reasoning. Since we assumed that the unrenormalized background electromagnetic field and the unrenormalized four-current density fulfill Maxwell's equations $\partial_{\mu}\mathcal{F}_0^{\mu\nu}(x)=\mathcal{J}_0^{\nu}(x)$ in deriving the Lagrangian density (\ref{Lb}) of SFQED and since $\mathcal{J}^{\mu}_0(x)$ has been renormalized as $\mathcal{J}^{\mu}_0(x) = \mathcal{J}^{\mu}(x)/\sqrt{Z_3}$, if we require that the physical (renormalized) background electromagnetic field $\mathcal{F}^{\mu\nu}(x)$ and four-current density $\mathcal{J}^{\mu}(x)$ also fulfill Maxwell's equations, then we have to impose that the background electromagnetic field is also renormalized as $\mathcal{A}^{\mu}_0(x) = \mathcal{A}^{\mu}(x)/\sqrt{Z_3}$. In conclusion, only based on the renormalization of vacuum QED, we can infer that in the case of the SFQED Lagrangian density (\ref{Lb}) the relations between the renormalized and unrenormalized quantities have to be \cite{Braun_1978,Brouder_2002} 
\begin{align}
\label{Fields_R} 
\psi(x)&=\frac{\psi_0(x)}{\sqrt{Z_2}}, & A^{\mu}(x)&=\frac{A_0^{\mu}(x)}{\sqrt{Z_3}}, & \mathcal{A}^{\mu}(x)&=\sqrt{Z_3}\mathcal{A}_0^{\mu}(x),\\ 
\label{Quantities_R}
m&=Z_2m_0-\delta m, & e&=\frac{Z_2\sqrt{Z_3}}{Z_1}e_0, & \rho&=\frac{\rho_0}{Z_3},
\end{align}
where all the renormalization constants are defined as in vacuum QED \cite{Itzykson_b_1980}. We remind that gauge invariance implies via the Ward identity that $Z_1=Z_2$ and that the gauge-fixing constant $\rho$ represents a redefinition of the gauge-fixing coefficient as it is an unphysical quantity, which does not undergo radiative corrections \cite{Itzykson_b_1980}. 

By introducing for future convenience the notation $\mathcal{L}_{B}$ for the Lagrangian density $\mathcal{L}_{B,0}$ expressed in terms of the renormalized fields and physical quantities (note that $\mathcal{L}_B=\mathcal{L}_{B,0}$), we can write
\begin{equation}
\label{L_B}
    \mathcal{L}_{B}=\mathcal{L}_{B,f}+\mathcal{L}_{B,i},
\end{equation}
where
\begin{equation}
\label{L_f_R}
\mathcal{L}_{B,f}=\bar{\psi}(i \hat{\partial}-e\hat{\mathcal{A}} -  m) \psi
- \frac{1}{4} F_{\mu \nu} F^{\mu \nu} - \frac{1}{2\rho}(\partial A)^2
\end{equation}
is the renormalized free Lagrangian density and
\begin{equation}
\label{L_i_R}
\begin{split}
\mathcal{L}_{B,i}&=-e\bar{\psi}\hat{A}\psi\\
&\quad+ (Z_2-1)\bar{\psi}i \hat{\partial}\psi-\delta m \bar{\psi}\psi-\left(\frac{Z_1}{Z_3}-1\right)e\bar{\psi}\hat{\mathcal{A}}\psi-\frac{Z_3-1}{4} F_{\mu \nu} F^{\mu \nu}-(Z_1-1)e\bar{\psi}\hat{A}\psi
\end{split}
\end{equation}
the renormalized interaction Lagrangian density, which includes the counterterms in the second line. Note that the first three counterterms provide corrections to fermion lines, the fourth to photon lines, and the last one to the vertex. The third counterterm is typical of SFQED and its role in the renormalization procedure will be studied below. Recalling the discussion on the Dirac four-current density at the beginning of Par. \ref{Tad_Tot}, we will not express the interaction Lagrangian density in terms of the commutators of the Dirac field and its Dirac conjugated field for the sake of notational simplicity. However, it will be understood that contractions between the Dirac field and its Dirac conjugated field at the same spacetime point have to be expressed via the symmetric time limit of the Feynman propagator.

Due to the somewhat unexpected difference in the renormalization of the background field $\mathcal{A}_0^{\mu}(x)$ as compared to the electromagnetic field operator $A_0^{\mu}(x)$, the question arises on how to renormalize the total electromagnetic-field operator $A_{T,0}^{\mu}(x)= \mathcal{A}_0^{\mu}(x) + A_0^{\mu}(x)$ or, equivalently, the total vacuum electromagnetic field $\mathcal{A}_{T,0}^{\mu}(x)=\mathcal{A}_0^{\mu}(x)+\mathcal{A}_{v,0}^{\mu}(x)$ (see Eqs. (\ref{A_phys}) and (\ref{A_phys-2})). From the definition
\begin{equation}
\left.
\mathcal{A}^\mu_{T,0}(x) = -    \frac{\delta W_0[0|J_0 + \mathcal{J}_0, \eta_0, \bar{\eta}_0]}{\delta J_{0,\mu}(x)  }
\right \vert_{J_0 = \eta_0 = \bar \eta_0 = 0 }
\end{equation}
and from the renormalization relation $W_0[0|J_0 + \mathcal{J}_0, \eta_0, \bar{\eta}_0] = W[0|J + \mathcal{J}, \eta, \bar{\eta}]$, we obtain
\begin{equation}
\left.
\mathcal{A}^\mu_{T,0}(x) = -  \sqrt{Z_3}\,  \frac{\delta W[0|J + \mathcal{J}, \eta, \bar{\eta}]}{\delta J_{\mu}(x)  }
\right \vert_{J = \eta = \bar \eta = 0 }
\end{equation}
and therefore that the total vacuum electromagnetic field has to renormalize as $\mathcal{A}_{T,0}^{\mu}(x)=\sqrt{Z_3}\mathcal{A}_T^{\mu}(x)$. By writing Eq. (\ref{W_0_J_W_0}) in terms of renormalized quantities:
\begin{equation}
\label{W_J_W}
W[0|J+\mathcal{J},\eta,\bar{\eta}]=W[\mathcal{A}|J,\eta,\bar{\eta}]+\frac{1}{Z_3}\int d^4 x\,\left[\frac{1}{4}\mathcal{F}_{\mu\nu}(x)\mathcal{F}^{\mu\nu}(x)-(J(x)\mathcal{A}(x))\right],
\end{equation}
with $W[\mathcal{A}|J,\eta,\bar{\eta}]=W_0[\mathcal{A}/\sqrt{Z_3}|J/\sqrt{Z_3},\eta/\sqrt{Z_2},\bar{\eta}/\sqrt{Z_2}]$, we obtain that
\begin{equation}
\label{A_T_Reno}
\mathcal{A}^\mu_{T}(x)= \left.
-    \frac{\delta W[\mathcal{A}|J, \eta, \bar{\eta}]}{\delta J_\mu(x)  }
\right \vert_{J = \eta = \bar \eta = 0 }+ \frac{1}{Z_3} \mathcal{A}^\mu(x).
\end{equation}
This equation shows that the SFQED one-point photon correlator, i.e., the tadpole, cannot be renormalized by the SFQED generator $W[\mathcal{A}|J, \eta, \bar{\eta}]$ alone. The same is true for the vacuum correlator because it is given by
\begin{equation}
W[\mathcal{A}|0,0,0]+\frac{1}{4Z_3}\int d^4 x\,\mathcal{F}_{\mu\nu}(x)\mathcal{F}^{\mu\nu}(x).
\end{equation}
Equation (\ref{W_J_W}) also implies that, apart from the mentioned vacuum correlator and the one-point photon correlator, all other renormalized correlators in SFQED are generated exclusively by $W[\mathcal{A}|J,\eta,\bar{\eta}]$.

The vacuum correlator will be studied elsewhere, whereas the one-point photon correlator will be investigated in detail below.

\subsection{The renormalization of the background  electromagnetic field and gauge invariance}
\label{Ren_A}
As we have already mentioned, the renormalization condition involving the background electromagnetic field $\mathcal{A}^{\mu}(x)$ is somewhat surprising, as it is different from that of the electromagnetic field operator. Here, we would like to further investigate this difference especially because one could question whether gauge invariance is endangered by the renormalization condition on the background electromagnetic field. 

In this paragraph, we first provide details on how technically the relative renormalization constant in Eq. (\ref{A_T_Reno}) arises and then we will show that gauge invariance is preserved.

Concerning the first point, we start from Eq. (\ref{A_phys}) and we follow a similar procedure as that provided in Ref. \cite{Braun_1972}. In fact, the electron propagator $G_0(x,y|\mathcal{A}_0)$ in Eq. (\ref{A_phys}) is the exact electron propagator stemming from all the contributions to the electron self energy including those featuring the tadpole itself. As we have discussed below Eq. (\ref{Eq_Prop}), we can write $G_0(x,y|\mathcal{A}_0)=G_{T,0}(x,y|\mathcal{A}_{T,0})$ and then (see Eqs. (\ref{A_v}) and (\ref{A_phys}))
\begin{equation}
\label{A_T}
\mathcal{A}^{\mu}_{T,0}(x)=\mathcal{A}_0^{\mu}(x)+ie_0\int d^4yD_0^{(0),\mu\nu}(x-y|0)\text{Tr}(\gamma_{\nu}G_{T,0}(y,y|\mathcal{A}_{T,0})).
\end{equation}
Now, we can expand the exact propagator $G_{T,0}(y,y|\mathcal{A}_{T,0})$ with respect to the total vacuum electromagnetic field $\mathcal{A}^{\mu}_{T,0}(x)$:
\begin{equation}
\begin{split}
\mathcal{A}^{\mu}_{T,0}(x)&=\mathcal{A}_0^{\mu}(x)+ie_0\sum_{n=0}^{\infty}\frac{1}{n!}\int d^4x_1\cdots d^4x_n\\
&\quad\times\int d^4yD_0^{(0),\mu\nu}(x-y|0)\text{Tr}\left(\gamma_{\nu}\frac{\delta^nG_{T,0}(y,y|0)}{\delta \mathcal{A}_{T,0}^{\nu_1}(x_1)\cdots \delta \mathcal{A}_{T,0}^{\nu_n}(x_n)}\right)\mathcal{A}_{T,0}^{\nu_1}(x_1)\cdots \mathcal{A}_{T,0}^{\nu_n}(x_n).
\end{split}
\end{equation}
The term with $n=0$ corresponds to the exact vacuum four-current density in the absence of the external field and it vanishes. The term with $n=1$, instead, is by definition given by \cite{Itzykson_b_1980}
\begin{equation}
\label{trace derivative Gtot}
ie_0\text{Tr}\left(\gamma_{\nu}\frac{\delta G_{T,0}(y,y|0)}{\delta \mathcal{A}_{T,0}^{\nu_1}(x_1)}\right)\equiv\Pi_{T,0,\nu\nu_1}(y-x_1|0),
\end{equation}
where $\Pi_{T,0,\nu\nu_1}(y-x_1|0)$ is the unrenormalized exact vacuum-polarization tensor in vacuum. Analogously the $n$-th coefficient of the expansion with $n>1$ can be shown to be given by
\begin{equation}
ie_0\text{Tr}\left(\gamma_{\nu}\frac{\delta^nG_{T,0}(y,y|0)}{\delta \mathcal{A}_{T,0}^{\nu_1}(x_1)\cdots \delta \mathcal{A}_{T,0}^{\nu_n}(x_n)}\right)\equiv \Pi_{T,0,\nu\nu_1\cdots\nu_n}(y,x_1,\ldots,x_n|0),
\end{equation}
where $\Pi_{T,0,\nu\nu_1\cdots\nu_n}(y,x_1,\ldots,x_n|0)$ is the (unrenormalized) exact, amputated, and one-particle irreducible (1PI) $(n+1)$-point photon correlation function in vacuum (recall that 1PI diagrams are connected diagrams that cannot be disconnected by cutting a single internal line \cite{Itzykson_b_1980}). In this way, we can write the expression of the total vacuum electromagnetic field $\mathcal{A}^{\mu}_{T,0}(x)$ as
\begin{equation}
\label{A_T_A_0_1}
\begin{split}
\mathcal{A}^{\mu}_{T,0}(x)&=\mathcal{A}_0^{\mu}(x)+\int d^4 x_1 d^4 yD_0^{(0),\mu\nu}(x-y|0)\Pi_{T,0,\nu\nu_1}(y-x_1|0)\mathcal{A}_{T,0}^{\nu_1}(x_1)+
\\&\sum_{n=2}^{\infty}\frac{1}{n!} \int d^4x_1\cdots d^4x_n d^4yD_0^{(0),\mu\nu}(x-y|0)\Pi_{T,0,\nu\nu_1\cdots\nu_n}(y,x_1,\ldots,x_n|0)\mathcal{A}_{T,0}^{\nu_1}(x_1)\cdots \mathcal{A}_{T,0}^{\nu_n}(x_n).
\end{split}
\end{equation}
In this expression the polarization operator is still superficially divergent, whereas all other correlation functions with $n> 1$ are superficially convergent. 

Now, we recall that, due to gauge invariance, the polarization operator in vacuum can be written as \cite{Itzykson_b_1980}
\begin{equation}
\label{Pi_0_x}
\begin{split}
\Pi_{T,0}^{\mu\nu}(x|0) &= \int \frac{d^4q}{(2\pi)^4}\, e^{-i(qx)}\check{\Pi}_{T,0}^{\mu\nu}(q|0)
= \int \frac{d^4q}{(2\pi)^4}\, e^{-i(qx)}(q^2\eta^{\mu\nu} - q^\mu q^\nu)\check{\Pi}_{T,0}(q^2|0)\\
&= -(\square\eta^{\mu\nu} - \partial^\mu\partial^\nu)\int \frac{d^4q}{(2\pi)^4}\, e^{-i(qx)}\check{\Pi}_{T,0}(q^2|0),
\end{split}
\end{equation}
where $\check{\Pi}_{T,0}^{\mu\nu}(q|0)=(q^2\eta^{\mu\nu} - q^\mu q^\nu)\check{\Pi}_{T,0}(q^2|0)$ is the polarization operator in momentum space. The renormalization of the polarization operator in vacuum is more easily carried out in momentum space \cite{Itzykson_b_1980} and
\begin{equation}
\label{Pi_0_q}
\check{\Pi}_{T,0}(q^2|0)=\check{\Pi}_{T,0}(q^2|0)-\check{\Pi}_{T,0}(0|0)+\check{\Pi}_{T,0}(0|0)=\frac{Z_3 - 1}{Z_3}+\frac{\Delta\check{\Pi}_T(q^2|0)}{Z_3},
\end{equation}
where $\Delta\check{\Pi}_T(q^2|0)$ is finite and $\check{\Pi}_{T,0}(0|0)=(Z_3 - 1)/Z_3$ \cite{Itzykson_b_1980}. Going back to the configuration space, we obtain
\begin{equation}
\label{Pi_0}
\Pi_{T,0}^{\mu\nu}(x|0) = -\frac{1}{Z_3}(\square\eta^{\mu\nu} - \partial^\mu\partial^\nu)\Delta\Pi_T(x|0)
- \frac{Z_3 - 1}{Z_3}(\square\eta^{\mu\nu} - \partial^\mu\partial^\nu)\delta^4(x),
\end{equation}
where $\Delta\Pi_T(x|0)$ is the inverse Fourier transform of the function $\Delta\check{\Pi}_T(q^2|0)$.

By using this expression of the polarization operator, we can rewrite Eq. (\ref{A_T_A_0_1}) as
\begin{equation}
\begin{split}\label{A_T-R-partV}
&\mathcal{A}^{\mu}_{T,0}(x)=\mathcal{A}_0^{\mu}(x)-\frac{1}{Z_3}\int d^4 x_1 d^4 yD_0^{(0),\mu\nu}(x-y|0)(\square_y\eta_{\nu\nu_1} - \partial_{y,\nu}\partial_{y,\nu_1})\Delta\Pi_T(y-x_1|0)\mathcal{A}_{T,0}^{\nu_1}(x_1)\\
&\quad-\frac{Z_3 - 1}{Z_3}\int d^4 x_1 d^4 yD_0^{(0),\mu\nu}(x-y|0)(\square_y\eta_{\nu\nu_1} - \partial_{y,\nu}\partial_{y,\nu_1})\delta^4(y-x_1)\mathcal{A}_{T,0}^{\nu_1}(x_1)+\sum_{n=2}^{\infty}\frac{1}{n!}\\
&\quad\times\int d^4x_1\cdots d^4x_n d^4yD_0^{(0),\mu\nu}(x-y|0)\Pi_{T,0,\nu\nu_1\cdots\nu_n}(y,x_1,\ldots,x_n|0)\mathcal{A}_{T,0}^{\nu_1}(x_1)\cdots \mathcal{A}_{T,0}^{\nu_n}(x_n).
\end{split}
\end{equation}
Due to the transverse structure of the polarization operator and of the $n$-point photon correlation functions with $n\ge 2$, the gauge-dependent terms in the free photon propagator do not contribute. Thus, by integrating by parts the integrals in $d^4y$ in the first two lines of this equation, we obtain (see Eq. (\ref{Free_Ph_Pr}))
\begin{equation}
\label{A_T_R}
\begin{split}
&\mathcal{A}^{\mu}_{T,0}(x)=\mathcal{A}_0^{\mu}(x)+\frac{1}{Z_3}\int d^4 x_1\Delta\Pi_T(x-x_1|0)\mathcal{A}_{T,0}^{\mu}(x_1)+\frac{Z_3 - 1}{Z_3}\mathcal{A}_{T,0}^{\mu}(x)\\
&\quad+\sum_{n=2}^{\infty}\frac{1}{n!}\int d^4x_1\cdots d^4x_n d^4yD_0^{(0),\mu\nu}(x-y|0)\Pi_{T,0,\nu\nu_1\cdots\nu_n}(y,x_1,\ldots,x_n|0)\mathcal{A}_{T,0}^{\nu_1}(x_1)\cdots \mathcal{A}_{T,0}^{\nu_n}(x_n).
\end{split}
\end{equation}
This equation is now suitable to introduce the remaining renormalized quantities according to
\begin{align}
\label{A_T_Ren}
\mathcal{A}_{T,0}^{\mu}(x)&=\sqrt{Z_3}\mathcal{A}_T^{\mu}(x),\\
\label{Pi_n}
\Pi_{T,0,\nu\nu_1\cdots\nu_n}(y,x_1,\ldots,x_n|0)&=Z_3^{-(n+1)/2}\Pi_{T,\nu\nu_1\cdots\nu_n}(y,x_1,\ldots,x_n|0), && \text{for $n\ge 2$}.
\end{align}
Notice that the function $D_0^{(0),\mu\nu}(x-y|0)$ does not undergo any renormalization as it is the explicit function given in Eq. (\ref{Free_Ph_Pr}). The resulting equation reads      
\begin{equation}
\label{A_T_R_2}
\begin{split}
&\mathcal{A}^{\mu}_{T}(x)=\sqrt{Z_3}\mathcal{A}_0^{\mu}(x)+\int d^4 x_1\Delta\Pi_T(x-x_1|0)\mathcal{A}_T^{\mu}(x_1)\\
&\quad+\sum_{n=2}^{\infty}\frac{1}{n!}\int d^4x_1\cdots d^4x_n d^4yD_0^{(0),\mu\nu}(x-y|0)\Pi_{T,\nu\nu_1\cdots\nu_n}(y,x_1,\ldots,x_n|0)\mathcal{A}_T^{\nu_1}(x_1)\cdots \mathcal{A}_T^{\nu_n}(x_n)
\end{split}
\end{equation}
and implies that, in order to obtain a finite relation among the renormalized quantities, the correct way of renormalizing the external field is according to $\sqrt{Z_3}\mathcal{A}_0^{\mu}(x)=\mathcal{A}^{\mu}(x)$ (as we have already noticed, the correlation functions $\Pi_{T,0,\nu\nu_1\cdots\nu_n}(y,x_1,\ldots,x_n|0)$ with $n > 1$ are superficially convergent).

Now, we come to the second point, i.e., why the above normalization procedure preserves gauge invariance. This is related to the fact that, according to Eq. (\ref{Eq_Prop_f}), the electromagnetic field effectively interacting with the charge is the total vacuum electromagnetic field and then gauge invariance is preserved because $e_0\mathcal{A}_{T,0}^{\mu}(x)=e\mathcal{A}_T^{\mu}(x)$. In addition, we would like to show here that this renormalization condition also guarantees that Eq. (\ref{Eq_Prop_f}) is fully renormalizable. In order to do this, we expand the unrenormalized mass operator $\Sigma_{T,0}(x,s|\mathcal{A}_{T,0})$ with respect to $\mathcal{A}_{T,0}^{\mu}(x)$. As it is clear from power counting (see also Ref. \cite{Brouder_2002}), only the vacuum term and the linear term with respect to $\mathcal{A}_{T,0}^{\mu}(x)$ are superficially divergent, whereas all higher-order terms are superficially convergent. By indicating as $\Delta\Sigma_T(x,s|\mathcal{A}_{T,0})$ the finite part of the mass operator (in the sense indicated below), it is (see Ref. \cite{Itzykson_b_1980} and also below)
\begin{equation}
\label{Sigma_ren}
\Sigma_{T,0}(x,s|\mathcal{A}_{T,0})=\delta^4(x-s)\left\{\frac{Z_2-1}{Z_2}[i\hat{\partial}_x-e_0\hat{\mathcal{A}}_{T,0}(x)]-\frac{\delta m}{Z_2}\right\}+\frac{\Delta\Sigma_T(x,s|\mathcal{A}_{T,0})}{Z_2}.
\end{equation}
Note that the function $\Delta\Sigma_T(x,s|\mathcal{A}_{T,0})$ is meant to be finite in the sense that, since $\Sigma_{T,0}(x,s|\mathcal{A}_{T,0})$ has the same diagrammatic structure as in the vacuum, it is renormalized as in vacuum. The only potentially divergent quantities still remaining are the occurrences of the unrenormalized field $\mathcal{A}^{\mu}_{T,0}(x)$, which always appears multiplied by $e_0$ (in all other occurrences detached from $\mathcal{A}^{\mu}_{T,0}(x)$ the renormalized electric charge appears). This is also the reason why we have indicated the finite part of the mass operator as dependent on $\mathcal{A}^{\mu}_{T,0}(x)$. By substituting Eq. (\ref{Sigma_ren}) in Eq. (\ref{Eq_Prop_f}), we obtain
\begin{equation}
\frac{1}{Z_2}[i\hat{\partial}_x-e_0\hat{\mathcal{A}}_{T,0}(x)-m]G_{T,0}(x,y|\mathcal{A}_{T,0})=\delta^4(x-y)+\frac{1}{Z_2}\int d^4s\,\Delta\Sigma_T(x,s|\mathcal{A}_{T,0})G_{T,0}(s,y|\mathcal{A}_{T,0}).
\end{equation}
At this point, we renormalize the electron propagator as in vacuum \cite{Itzykson_b_1980}, i.e., as $G_{T,0}(x,y|\mathcal{A}_{T,0})=Z_2G_T(x,y|\mathcal{A}_{T,0})$, where we have kept the dependence on $\mathcal{A}^{\mu}_{T,0}(x)$ for the same reason as in $\Delta\Sigma_T(x,s|\mathcal{A}_{T,0})$ and we have
\begin{equation}
[i\hat{\partial}_x-e_0\hat{\mathcal{A}}_{T,0}(x)-m]G_T(x,y|\mathcal{A}_{T,0})=\delta^4(x-y)+\int d^4s\,\Delta\Sigma_T(x,s|\mathcal{A}_{T,0})G_T(s,y|\mathcal{A}_{T,0}).
\end{equation}
Finally, this equation shows that by only renormalizing the total vacuum electromagnetic field as the quantum electromagnetic field, i.e., such that $e_0\mathcal{A}^{\mu}_{T,0}(x)=e\mathcal{A}^{\mu}_T(x)$, all the divergences of this equation cancel out and we finally obtain the finite equation
\begin{equation}
[i\hat{\partial}_x-e\hat{\mathcal{A}}_T(x)-m]G_T(x,y|\mathcal{A}_T)=\delta^4(x-y)+\int d^4s\,\Delta\Sigma_T(x,s|\mathcal{A}_T)G_T(s,y|\mathcal{A}_T).
\end{equation}

\subsection{Collins' and Weinberg's approaches to renormalization of SFQED}
\label{sub-sec Collins and Weinberg}
The monographs \cite{Collins_b_1984,Weinberg_b_1_2005} present alternative ways of introducing SFQED and of deriving the renormalized SFQED Lagrangian density starting from the vacuum QED Lagrangian density. The aim of this paragraph is to show that these alternative ways of renormalizing SFQED are in agreement with the Lagrangian density $\mathcal{L}_B$ in Eq. (\ref{L_B}). Below, we adapt the notation of Refs. \cite{Collins_b_1984,Weinberg_b_1_2005} to ours.

The procedures followed in Refs. \cite{Collins_b_1984,Weinberg_b_1_2005} are sufficiently different from ours to deserve some clarifications. In particular, both monographs directly introduce the renormalized background field without mentioning the unrenormalized one. 

In Ref. \cite{Collins_b_1984} Collins first introduces the renormalized Lagrangian density of QED in the presence of the external four-current density $\mathcal{J}^{\mu}(x)$:
\begin{equation}
\label{L_C}
\begin{split}
\mathcal{L}_C&=\bar{\psi}(i \hat{\partial} -  m ) \psi
- \frac{1}{4} F_{C,\mu \nu} F_C^{\mu \nu} - \frac{1}{2\rho}(\partial A_C)^2-(\mathcal{J}A_C)-e\bar{\psi}\hat{A}_C\psi\\
&\quad + (Z_2-1)\bar{\psi}i \hat{\partial}\psi-\delta m \bar{\psi}\psi-\frac{Z_3-1}{4} F_{C,\mu \nu} F_C^{\mu \nu}-(Z_1-1)e\bar{\psi}\hat{A}_C\psi,
\end{split}
\end{equation}
where we have purposely added the index $C$ (standing for ``Collins'') to the renormalized electromagnetic field for a reason which will be clear below.

The renormalized background field $\mathcal{A}^{\mu}(x)$ is introduced by shifting $A_C^{\mu}(x)\to A_C^{\mu}(x)+\mathcal{A}^{\mu}(x)$ and assuming that it satisfies the Lorenz-gauge condition $(\partial\mathcal{A}(x))=0$ and Maxwell's equations $\square \mathcal{A}^{\mu}(x)=\mathcal{J}^{\mu}(x)$. After ignoring the non-dynamical terms, we obtain Collins' form $\mathcal{L}_{B,C}$ of the SFQED Lagrangian density
\begin{equation}
\label{L_B_C}
\begin{split}
\mathcal{L}_{B,C}&=\bar{\psi}(i \hat{\partial}-e\hat{\mathcal{A}} -  m ) \psi
- \frac{1}{4} F_{C,\mu \nu} F_C^{\mu \nu} - \frac{1}{2\rho}(\partial A_C)^2-e\bar{\psi}\hat{A}_C\psi\\
&\quad + (Z_2-1)\bar{\psi}i \hat{\partial}\psi-\delta m \bar{\psi}\psi-\frac{Z_3-1}{4} F_{C,\mu \nu} F_C^{\mu \nu}-(Z_1-1)e\bar{\psi}\hat{A}_C\psi\\
&\quad+(Z_3-1)(\mathcal{J}A_C)-(Z_1-1)e\bar{\psi}\hat{\mathcal{A}}\psi.
\end{split}
\end{equation}
The equivalence with the Lagrangian density $\mathcal{L}_B$ is obtained once one observes that the field $A^{\mu}(x)$ satisfies the equation
\begin{equation}
Z_3\left[\square  \eta^{\mu \nu} - \left(1-\frac{1}{\rho} \right) \partial^\mu \, \partial^\nu\right]A_{\nu}=Z_1e\bar{\psi}\gamma^{\mu}\psi,
\end{equation}
whereas the field $A_C^{\mu}(x)$ satisfies the equation
\begin{equation}
\label{Eq_M_A_C}
Z_3\left[\square  \eta^{\mu \nu} - \left(1-\frac{1}{\rho} \right) \partial^\mu \, \partial^\nu\right]A_{C,\nu}=Z_1e\bar{\psi}\gamma^{\mu}\psi-(Z_3-1)\left[\square  \eta^{\mu \nu} - \left(1-\frac{1}{\rho} \right) \partial^\mu \, \partial^\nu\right]\mathcal{A}_{\nu},
\end{equation}
where we have used the fact that $(\partial\mathcal{A}(x))=0$. Thus, the two fields $A^{\mu}(x)$ and $A_C^{\mu}(x)$ are not identical but
\begin{equation}
\label{A_C}
A^{\mu}(x)=A_C^{\mu}(x)+\left(1-\frac{1}{Z_3}\right)\mathcal{A}^{\mu}(x).
\end{equation}
Then, after one performs this substitution in the Lagrangian density $\mathcal{L}_{B,C}$, one can easily obtain the Lagrangian density $\mathcal{L}_B$ up to non-dynamical terms. 

We now pass to Weinberg's approach in Ref. \cite{Weinberg_b_1_2005}, where he starts from the renormalized Lagrangian density 
\begin{equation}
\label{L_W}
\begin{split}
\mathcal{L}_W&=\bar{\psi}(i \hat{\partial} -  m ) \psi
- \frac{1}{4} F_{W,\mu \nu} F_W^{\mu \nu} - \frac{1}{2\rho}(\partial A_W)^2-e\bar{\psi}\hat{A}_W\psi\\
&\quad + (Z_2-1)\bar{\psi}i \hat{\partial}\psi-\delta m \bar{\psi}\psi-\frac{Z_3-1}{4} F_{W,\mu \nu} F_W^{\mu \nu}-(Z_1-1)e\bar{\psi}\hat{A}_W\psi,
\end{split}
\end{equation}
with the index $W$ (standing for ``Weinberg'') being added for notational convenience. Weinberg introduces the background field by making the substitution $A_W^{\mu}(x)\to A_W^{\mu}(x)+\mathcal{A}^{\mu}(x)$ only in the interaction terms which include the counterterms. The resulting Lagrangian density (apart from a non-dynamical term)
\begin{equation}
\label{L_B_W}
\begin{split}
\mathcal{L}_{B,W}&=\bar{\psi}(i \hat{\partial}-e\hat{\mathcal{A}} -  m ) \psi
- \frac{1}{4} F_{W,\mu \nu} F_W^{\mu \nu} - \frac{1}{2\rho}(\partial A_W)^2-e\bar{\psi}\hat{A}_W\psi\\
&\quad + (Z_2-1)\bar{\psi}i \hat{\partial}\psi-\delta m \bar{\psi}\psi-\frac{Z_3-1}{4} F_{W,\mu \nu} F_W^{\mu \nu}-(Z_1-1)e\bar{\psi}\hat{A}_W\psi\\
&\quad-\frac{Z_3-1}{2} \mathcal{F}_{\mu \nu} F_W^{\mu \nu}-(Z_1-1)e\bar{\psi}\hat{\mathcal{A}}\psi,
\end{split}
\end{equation}
is easily found to coincide with Collins' Lagrangian density $\mathcal{L}_{B,C}$ by identifying $A_W^{\mu}(x)=A_C^{\mu}(x)$ and after using the equation of motion $\square \mathcal{A}^{\mu}(x)=\mathcal{J}^{\mu}(x)$ of $\mathcal{A}^{\mu}(x)$ to obtain the term proportional to the four-current density $\mathcal{J}^{\mu}(x)$ in Eq. (\ref{L_B_C}). At that point, one can proceed as before, using the field redefinition in Eq. \eqref{A_C}, such that also Weinberg's approach ultimately leads to the Lagrangian density $\mathcal{L}_B$ in Eq. (\ref{L_B}).

According to the above results, one can use either our Lagrangian density $\mathcal{L}_B$ or Collins (Weinberg's) Lagrangian density in Eq. (\ref{L_B_C}) (Eq. (\ref{L_B_W})), without even referring to the unrenormalized background field and four-current density. Since Collin's and Weinberg's approaches are essentially identical, we refer only to our approach and to Collin's one and we observe that their equivalence can also be seen by computing the generating functionals $Z_C[\mathcal{A}|J, \eta, \bar{\eta}]$ and $Z[\mathcal{A}|J, \eta, \bar{\eta}]$ of the renormalized Green’s functions. In fact, starting from Eqs. (\ref{L_f_R}), (\ref{L_i_R}), and (\ref{L_B_C}) and by shifting the integrated field $A_C^{\mu}(x)$ in $Z_C[\mathcal{A}|J, \eta, \bar{\eta}]$ according to Eq. (\ref{A_C}), one can easily show that
\begin{equation} 
\label{Z_C_Z}
		Z_C[\mathcal{A}|J, \eta, \bar{\eta}]  =  Z[\mathcal{A}|J, \eta, \bar{\eta}] 
		 \, e^{i\frac{Z_3-1}{Z_3}\int d^4x\, \left[\frac{Z_3-1}{4}\F_{\mu \nu} \F^{\mu \nu}  +(J\A)\right]}.
	\end{equation}
Now, by introducing the generator $W_C[\mathcal{A}|J, \eta, \bar{\eta}]=-i\log(Z_C[\mathcal{A}|J, \eta, \bar{\eta}])$ of the connected correlators in Collins' approach, one sees that all correlators with two or more external lines coincide with those in our approach, whereas the vacuum-vacuum transition amplitudes differ by an inconsequential constant (which matters for the renormalization, though) and the one-particle photon correlators differ according to Eq. (\ref{A_C}). 

Now, by comparing Eq. (\ref{Z_C_Z}) in terms of the generators of the connected correlators with Eq. (\ref{W_J_W}), we also obtain
\begin{equation}
W[0|J+\mathcal{J},\eta,\bar{\eta}]=W_C[\mathcal{A}|J,\eta,\bar{\eta}]-\int d^4 x\,\left[\frac{Z_3-2}{4}\mathcal{F}_{\mu\nu}(x)\mathcal{F}^{\mu\nu}(x)+(J(x)\mathcal{A}(x))\right].
\end{equation}
Apart from showing the explicit equivalence between Collins' approach and the one based on vacuum QED plus the background four-current density, this equation allows for deriving the relation between the total vacuum electromagnetic field $\mathcal{A}^\mu_{T}(x)$ and the tadpole in Collins' approach: 
\begin{equation}
\label{A_T_C}
\mathcal{A}^\mu_{T}(x)= -\left.
\frac{\delta W_C[\mathcal{A}|J, \eta, \bar{\eta}]}{\delta J_\mu(x)  }
\right \vert_{J = \eta = \bar \eta = 0 }+  \mathcal{A}^\mu(x). 
\end{equation}
This equality shows explicitly that, unlike the tadpole in the theory based on the SFQED Lagrangian density $\mathcal{L}_B$ (see Eq. (\ref{A_T_Reno}) and Par. \ref{Ren_A}), the tadpole in Collins' theory can be renormalized on its own. The reason is that, unlike the Lagrangian density $\mathcal{L}_B$ (see Eq. (\ref{L_i_R})), Collins' Lagrangian density features a counterterm which explicitly renormalizes the tadpole, namely the first one in the last line of Eq. (\ref{L_B_C}). As we have mentioned, all these considerations also apply to Weinberg's Lagrangian density whose counterterm renormalizing the tadpole is the first one in the last line of Eq. (\ref{L_B_W}).

Although the above proof already shows the equivalence between our approach and that of Collins and Weinberg, it is instructive to investigate the equivalence also at the level of the unrenormalized quantities. Again, we only refer to Collins' Lagrangian density and we observe that he carries out the shift on the electromagnetic field to include the background electromagnetic field at the level of the renormalized fields. Thus, one concludes that in Collins' approach the renormalization of the background field is actually carried out as for the quantum electromagnetic field, i.e., $\mathcal{A}_{0,C}^{\mu}(x)=\sqrt{Z_3}\mathcal{A}^{\mu}(x)$. In turn, the relation in Eq. (\ref{A_C}) among the renormalized fields implies that $A_0^{\mu}(x)+\mathcal{A}_0^{\mu}(x)=A_{0,C}^{\mu}(x)+\mathcal{A}_{0,C}^{\mu}(x)$. Then, by using the equation $\mathcal{J}_0^{\mu}(x)=\mathcal{J}^{\mu}(x)/\sqrt{Z_3}$, one sees that the unrenormalized Lagrangian density in Collins' approach is exactly Eq. (\ref{L_B_0}) with $A_0^{\mu}(x)\to A_{0,C}^{\mu}(x)$ and $\mathcal{A}_0^{\mu}(x)\to \mathcal{A}_{0,C}^{\nu}(x)$. In other words, starting from the same Lagrangian density Eq. (\ref{L_B_0}), it is important that the background four-current density renormalizes as $\mathcal{J}_0^{\mu}(x)=\mathcal{J}^{\mu}(x)/\sqrt{Z_3}$ but one can then renormalize the background field either as we have done or as Collins did. Since in both approaches the renormalized background electromagnetic field and four-current density have to satisfy Maxwell's equations $\partial_{\mu}\mathcal{F}^{\mu\nu}(x)=\mathcal{J}^{\nu}(x)$, this implies that according to our renormalization condition $\mathcal{A}_0^{\mu}(x)=\mathcal{A}^{\mu}(x)/\sqrt{Z_3}$, it is $\partial_{\mu} \mathcal{F}_0^{\mu\nu}(x)=\mathcal{J}_0^{\nu}(x)$, i.e., also the unrenormalized background electromagnetic field and four-current density satisfy Maxwell's equations, whereas within Collins' approach it is $\partial_{\mu} \mathcal{F}_{C,0}^{\mu\nu}(x)= Z_3 \mathcal{J}_0^{\mu}(x)$. In conclusion, the two approaches share the same unrenormalized Lagrangian density in Eq. (\ref{L_B_0}) (with $A_0^{\mu}(x)\to A_{0,C}^{\mu}(x)$ and $\mathcal{A}_0^{\mu}(x)\to \mathcal{A}_{0,C}^{\mu}(x)$ in Collins' approach) but then, if one imposes $\partial_{\mu} \mathcal{F}_0^{\mu\nu}(x)=\mathcal{J}_0^{\nu}(x)$, as we did, one arrives to Eq. (\ref{Lb}), whereas if one imposes $\partial_{\mu} \mathcal{F}_{C,0}^{\mu\nu}(x)= Z_3 \mathcal{J}_0^{\nu}(x)$, as in Collins' approach, one would have an extra term $(Z_3-1)(\mathcal{J}_0A_{C,0})$, such that after renormalization Eq. (\ref{L_C}) is recovered.

According to the above discussion, the renormalization of the background four-current density seems to be more primary than that of the background field. Also, the three renormalized SFQED Lagrangian densities $\mathcal{L}_B$ (see Eq. (\ref{L_B})), $\mathcal{L}_{B,C}$ (see Eq. (\ref{L_B_C})), and $\mathcal{L}_{B,W}$ (see Eq. (\ref{L_B_W})) are physically equivalent. However, if one wants to introduce the unrenormalized background electromagnetic field starting from the unrenormalized vacuum QED Lagrangian density in Eq. (\ref{L_V}), then one has to keep in mind that, by using the unrenormalized SFQED Lagrangian density in Eq. (\ref{Lb}), one has already assumed that the unrenormalized background electromagnetic field and the unrenormalized background four-current density are related by Maxwell's equations, and then that the background electromagnetic field is renormalized as a charge.

Accounting for the equivalence of all the discussed approaches, we continue using the Lagrangian density $\mathcal{L}_B$ and the renormalization condition $\mathcal{A}^{\mu}(x)=\sqrt{Z_3}\mathcal{A}_0^{\mu}(x)$ for the background electromagnetic field.

\section{Determination of the renormalization constants in perturbation theory}\label{sec-ren-ct}
Once we have ascertained that the renormalized Lagrangian density $\mathcal{L}_B$ of SFQED is given by the sum of the two Lagrangian densities in Eqs. (\ref{L_f_R}) and (\ref{L_i_R}), the next task is to give a prescription to determine the renormalization constants $\delta m$, $Z_1$, $Z_2$, and $Z_3$ (recall that gauge invariance implies that $Z_1=Z_2$).

We first remind that the only superficially divergent connected diagrams in vacuum QED are \cite{Itzykson_b_1980}:
\begin{enumerate}
\item diagrams with no external lines (vacuum diagrams), whose overall sum reduces to a global (divergent) phase and which, as such, do not contribute to any transition probability;
\item diagrams with one-photon external line, which actually vanish due to charge-parity conservation in QED (Furry theorem);
\item diagrams with two fermion external lines;
\item diagrams with two photon external lines;
\item diagrams with three photon external lines, which actually vanish due to charge-parity conservation in QED (Furry theorem);
\item diagrams with two fermion and one photon external lines;
\item diagrams with four photon external lines, which actually converge due to gauge invariance.
\end{enumerate}
In this context it is useful to refer to the 1PI diagrams corresponding to the three classes of non-vanishing and physically relevant superficially divergent connected diagrams in the above list. The sum of all 1PI diagrams with two fermion external lines is known as mass operator, the sum of all 1PI diagrams with two photon external lines is known as polarization operator, and the sum of all 1PI diagrams with two fermion and one photon external lines is known as vertex correction. These are the diagrams that one ultimately needs to renormalize in QED in vacuum.

Moving now to the superficially divergent connected diagrams in SFQED, a few additional observations are in order. First, the background electromagnetic field may render the vacuum unstable and vacuum diagrams may contribute to the so-called vacuum persistence probability amplitude in Eq. \eqref{persistence-apt}, i.e., the probability amplitude that the vacuum remains vacuum \cite{Fradkin_b_1991}. As we have already mentioned, we will investigate the vacuum diagrams and how they can be renormalized without needing a new renormalization constant elsewhere. Second, the diagrams with one external photon line, whose overall sum is the tadpole diagram in Fig. \ref{Tadpole_Exact}, do not vanish, which is related to the fact that the electron propagators in SFQED feature an arbitrary number of insertions of background-field lines. Third, for the same reason also the diagrams with three photon external lines do not vanish as in vacuum. As we will see in detail and as it is already clear from Eq. (\ref{A_T_Reno}), the tadpole diagrams do not feature a genuinely new divergence as compared to vacuum QED and the diagrams with three external photon lines are actually convergent, which is in agreement with the fact that the number of renormalization constants in SFQED is the same as in vacuum. Nevertheless, the SFQED Lagrangian density $\mathcal{L}_B$ features an additional counterterm as compared to the vacuum QED, which is the third term in the second line of Eq. (\ref{L_i_R}).

In order to analyze the divergences of the 1PI superficially divergent diagrams in SFQED, we first observe that the presence of the external electromagnetic field does not introduce new divergences. We can prove this statement for the one-loop, 1PI superficially divergent diagrams because, as in vacuum QED, the divergences of higher-loop diagrams can be then removed by means of the Bogoliubov-Parasiuk-Hepp-Zimmermann (BPHZ) inductive method \cite{Itzykson_b_1980}. The only point to be clarified is how to treat the tadpoles, which identically vanish in vacuum QED and which will be discussed below.

\subsection{Vertex correction}

\begin{figure}
\includegraphics[width=3cm]{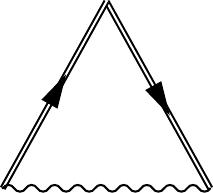}
\caption{One-loop vertex correction in an external field.}
\label{VC_no_Legs}
\end{figure}

\begin{figure}
\includegraphics[width=15cm]{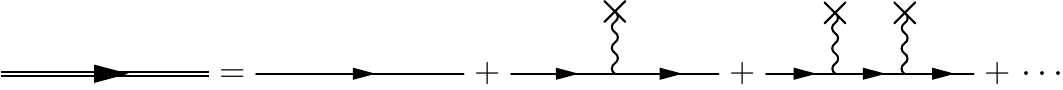}
\caption{Dressed electron propagator in a background field and at the tree-level in the interaction between the Dirac field and the quantum radiation field graphically expanded in a series with respect to the background field. The photon lines ending with a crossed vertex indicate the external field.}
\label{Dressed_Prop}
\end{figure}

The fact that the presence of the background electromagnetic field does not introduce new divergences is particularly clear in the case of the vertex correction (see Fig. \ref{VC_no_Legs}).
In fact, by imagining to expand the corresponding amplitude in SFQED in powers of the external field, it contains a logarithmically divergent vacuum contribution plus all other field-dependent contributions, which converge by power counting, as it can be easily ascertained by adding one or more background-field vertices to any of the two electron propagators in the vertex correction in vacuum (see also Fig. \ref{Dressed_Prop}).
Thus, the constant $Z_1$ can be effectively determined as in vacuum because the difference between the strong-field vertex correction and the vacuum vertex correction is finite. Enforcing the renormalization condition on $Z_1$ as in vacuum operationally corresponds to define the physical electric charge via a collision experiment carried out in vacuum, which is physically reasonable. The counterterm removing the divergence of the vertex correction is the last one in the second line of Eq. (\ref{L_i_R}). By diagrammatically indicating the counterterm as an additional vertex with a circled cross, the corresponding Feynman rule in coordinate space reads
\begin{equation}
\label{vertex_CT}
    \begin{tikzpicture}[baseline={([yshift=-0.5ex]current bounding box.center)}]
        \begin{feynman}
            \vertex (a1);
            \vertex [right=1.8cm of a1] (a2);
            \vertex (a3) at (0.9,0.9);
            \node[crossed dot] (c) at (0.9,0);
            \diagram*{
                (c) -- [doublefermion] (a1), (a2) -- [doublefermion] (c),
                (c) -- [photon] (a3)
            };
        \end{feynman}
    \end{tikzpicture}
    \quad \rightarrow \quad -ie(Z_1 -1)\gamma^\mu,
\end{equation}
with an additional four-dimensional integral.

\subsection{Polarization operator}
\begin{figure}
\includegraphics[width=3cm]{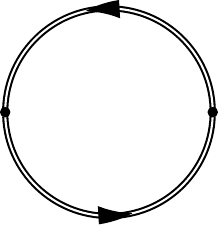}
\caption{One-loop polarization operator in an external field.}
\label{PO_no_Legs}
\end{figure}

The same reasoning can be  applied to the one-loop polarization operator (see Fig. \ref{PO_no_Legs})
but here, after expanding again the two tree-level dressed electron propagators in powers of the external field, one has to exploit the charge-parity symmetry and the gauge invariance of QED to arrive to the same conclusion as for the vertex correction. In fact, the contribution linear in the external field vanishes due to Furry theorem and the quadratic one is finite due to gauge invariance. This allows one to conclude that the difference between the vacuum polarization amplitude in a background field and in the vacuum is finite, and then that the constant $Z_3$ can be determined as in vacuum. With the same reasoning one can prove that the one-loop 1PI diagram with three external photons lines in SFQED is finite. The counterterm removing the divergence of the polarization operator is the fourth one in the second line of Eq. (\ref{L_i_R}) and the corresponding Feynman rule in coordinate space reads
\begin{equation}
\label{photon_CT}
    \begin{tikzpicture}[baseline={([yshift=-0.5ex]current bounding box.center)}]
        \begin{feynman}
            \vertex (a1);
            \vertex [right=1.8cm of a1] (a2);
            \node[crossed dot] (c) at (0.9,0);
            \diagram*{
                (a1) -- [photon] (c) -- [photon] (a2),
            };
        \end{feynman}
    \end{tikzpicture}
    \quad \rightarrow \quad 
    i (Z_3 -1)(\square\eta^{\mu\nu} - \partial^\mu\partial^\nu),
\end{equation}
with an additional four-dimensional integral.

\subsection{Mass operator and tadpole}
Now, we turn to the discussion of the one-loop mass operator and then of the one-loop tadpole, which, as we have mentioned, can be interpreted as corrections to the electron self energy. Indeed, the counterterms used to remove the corresponding divergences are the first three in the second line of Eq. (\ref{L_i_R}), containing only two fermion fields. Before starting a detailed analysis of these counterterms, we write the corresponding Feynman rule. This is the only case where the counterterms are structurally different from those in vacuum and therefore we use a crossed square to indicate them in a Feynman diagram. Looking at the first three terms in the second line of Eq. (\ref{L_i_R}), it is 
\begin{equation}
\label{CT_fermion_0}
    \begin{tikzpicture}[baseline={([yshift=-0.5ex]current bounding box.center)}]
        \begin{feynman}
            \vertex (a1);
            \vertex [right=1.8cm of a1] (a2);
            \node[crossed square] (c) at (0.9,0);
            \diagram*{
                (a2) -- [doublefermion] (c) -- [doublefermion] (a1),
            };
        \end{feynman}
    \end{tikzpicture}
    \quad \rightarrow \quad i\left[(Z_2 -1)i\hat{\partial}-\delta m -\left(\frac{Z_1}{Z_3} -1\right) e \hat{\mathcal{A}}(x)\right],
\end{equation}
with the additional four-dimensional integral.

\subsubsection{Mass operator}
\begin{figure}
\includegraphics[width=3cm]{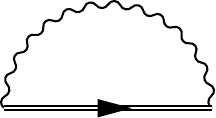}
\caption{One-loop mass operator in an external field.}
\label{MO_no_Legs}
\end{figure}

\begin{figure}
\includegraphics[width=3cm]{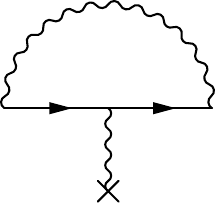}
\caption{First-order contribution in the external field to the one-loop mass operator. The photon lines ending with a crossed vertex indicate the external field.}
\label{MO_no_Legs_1}
\end{figure}

Let us now first consider the one-loop mass operator in SFQED (see Fig. \ref{MO_no_Legs}).
By expanding the dressed electron propagator in powers of the external field, we recognize that, apart from the vacuum contribution, also the contribution linear in the field is divergent (see Fig. \ref{MO_no_Legs_1}). 
While the counterterms proportional to $Z_2-1$ and to $\delta m$ in Eq. (\ref{L_i_R}), corresponding to the first two terms in Eq. (\ref{CT_fermion_0}), can be used to remove the divergences of the vacuum contribution, an additional term is required to remove the divergence linear in the background field. Since the resulting divergence is that of a  vertex correction with the external photon line being replaced by the background field (see Fig. \ref{MO_no_Legs_1}), it is convenient to write the third counterterm in the second line of Eq. (\ref{L_i_R}) as
\begin{equation}
\label{Counter_SFQED}
-\left(\frac{Z_1}{Z_3}-1\right)e\bar{\psi}(x)\hat{\mathcal{A}}(x)\psi(x)=-(Z_1-1)e\bar{\psi}(x)\hat{\mathcal{A}}(x)\psi(x)+\frac{Z_1}{Z_3}(Z_3-1)e\bar{\psi}(x)\hat{\mathcal{A}}(x)\psi(x),
\end{equation}
because the counterterm $-(Z_1-1)e\bar{\psi}(x)\hat{\mathcal{A}}(x)\psi(x)$ has exactly the right structure and proportionality constant to remove that type of vertex-correction divergence in the mass operator. We can exploit the identity $Z_1=Z_2$ resulting from gauge invariance \cite{Itzykson_b_1980}, to combine this counterterm with the derivative one such that the Feynman rule (\ref{CT_fermion_0}) can be written as
\begin{equation}
\label{CT_fermion_1}
    \begin{tikzpicture}[baseline={([yshift=-0.5ex]current bounding box.center)}]
        \begin{feynman}
            \vertex (a1);
            \vertex [right=1.8cm of a1] (a2);
            \node[crossed square] (c) at (0.9,0);
            \diagram*{
                (a2) -- [doublefermion] (c) -- [doublefermion] (a1),
            };
        \end{feynman}
\end{tikzpicture}
=
\begin{tikzpicture}[baseline={([yshift=-0.5ex]current bounding box.center)}]
        \begin{feynman}
            \vertex (a1);
            \vertex [right=1.8cm of a1] (a2);
            \node[crossed dot] (c) at (0.9,0);
            \diagram*{
                (a2) -- [doublefermion] (c) -- [doublefermion] (a1),
            };
        \end{feynman}
\end{tikzpicture}
+ 
\begin{tikzpicture}[baseline={([yshift=-2.5ex]current bounding box.center)}]
        \begin{feynman}
            \vertex (a1);
            \vertex [right=1.8cm of a1] (a2);
            \vertex (e) at (0.9,0.9) ;
            \node[crossed triangle down] (c) at (0.9,0) { $\times$};
            \node[inner sep=0pt] (d) at (0.9, 0.9) {$\times$};
            \diagram*{
                (a2) -- [doublefermion] (c) -- [doublefermion] (a1),
                (c) -- [photon] (e)
            };
        \end{feynman}      
\end{tikzpicture},
\end{equation}
with
\begin{align}
\label{CT_fermion_split_1}
    &\begin{tikzpicture}[baseline={([yshift=-0.5ex]current bounding box.center)}]
        \begin{feynman}
            \vertex (a1);
            \vertex [right=1.8cm of a1] (a2);
            \node[crossed dot] (c) at (0.9,0);
            \diagram*{
                (a2) -- [doublefermion] (c) -- [doublefermion] (a1),
            };
        \end{feynman}
    \end{tikzpicture}
    \quad \rightarrow \quad i\left\{(Z_2 -1)[i\hat{\partial}-e\hat{\mathcal{A}}(x)]-\delta m \right\},
    \\ 
    \label{CT_fermion_split_2}
    &
        \begin{tikzpicture}[baseline={([yshift=-0.5ex]current bounding box.center)}]
        \begin{feynman}
            \vertex (a1);
            \vertex [right=1.8cm of a1] (a2);
            \vertex (e) at (0.9,0.9) ;
            \node[crossed triangle down] (c) at (0.9,0) { $\times$};
            \node[inner sep=0pt] (d) at (0.9, 0.9) {$\times$};
            \diagram*{
                (a2) -- [doublefermion] (c) -- [doublefermion] (a1),
                (c) -- [photon] (e)
            };
        \end{feynman}
    \end{tikzpicture}
    \quad \rightarrow \quad i\frac{Z_1}{Z_3}(Z_3 -1)\, e \hat{\mathcal{A}}(x).
\end{align}
The above analysis allows us to conclude that the first term renormalizes the mass operator. We used the crossed circle as the other counterterms analogous to the vacuum ones because the resulting counterterm $(Z_2-1)\bar{\psi}(x)[i \hat{\partial}-e\hat{\mathcal{A}}(x)]\psi(x) -\delta m\bar{\psi}(x)\psi(x)\sim [(Z_2-1)m-\delta m]\bar{\psi}(x)\psi(x)$ precisely corresponds to the vacuum one although it features the one-particle gauge-covariant four-momentum-operator in the background field (recall that in the $S$-matrix the fields are expanded in solutions of the Dirac equation in the presence of the background field $\mathcal{A}^{\mu}(x)$). Analogously as for the electron charge, the renormalization of the electron mass as in vacuum implies that the physical mass corresponds to the pole of the exact electron propagator in vacuum. It is known that in the presence of an external field, the electron mass can effectively undergo corrections as in the case of a background plane-wave field due to the interaction between the electron's intrinsic magnetic moment and the magnetic field of the plane wave (see Refs. \cite{Ritus_1970,Baier_1971} for the constant-crossed field case and Ref. \cite{Di_Piazza_2021_e} for the general plane-wave field case). Something analogous can be concluded about the renormalization constant $Z_2$, which, however, is not measurable.

\subsubsection{Tadpole}

Finally, we discuss the tadpole diagram, which is also divergent in SFQED (see Fig. \ref{Tadpole_Exact} and Eq. (\ref{A_T_Reno})). Indeed, its divergence is related to the second counterterm in Eq. (\ref{Counter_SFQED}) and then to the second Feynman rule in Eq. (\ref{CT_fermion_1}) (see Eq. (\ref{CT_fermion_split_2})), which is proportional to the background field. 

It is instructive at this point to make a comparison with the Lagrangian density of either Collins or Weinberg (see Eqs. (\ref{L_B_C}) and (\ref{L_B_W})). In fact, the analysis of the counterterms carried out so far, including the one renormalizing the mass operator, also applies to Collins' and Weinberg's Lagrangian densities. However, unlike our interaction Lagrangian density in Eq. (\ref{L_i_R}), Collins' and Weinberg's Lagrangian densities feature a counterterm for the tadpole, i.e., the first term in the last lines of Eqs. (\ref{L_B_C}) and (\ref{L_B_W}), which is, in fact, linear in the quantum radiation field. We have already observed that these two counterterms are equivalent up to a total derivative and we refer to Weinberg's Lagrangian density to conclude that the Feynman rule for that counterterm is 
\begin{equation}
\label{CT_Tadpole}
       \begin{tikzpicture}[baseline={([yshift=-0.5ex]current bounding box.center)}]
        \begin{feynman}
            \node[crossed triangle down] (c) at (0.9,1.2) { $\times$};
            \node (e) at (0.9, 0.) ;
            \diagram*{
                (c) -- [photon] (e)
            };
        \end{feynman}
    \end{tikzpicture}
    \quad \rightarrow \quad -(Z_3 -1) \mathcal{A}^{\mu}(x),
\end{equation}
where we have also included the photon propagator, corresponding to the photon line. Note that if one attaches this photon line to a vertex, one needs to contract $-(Z_3 -1) \mathcal{A}^{\mu}(x)$ with $-ie\gamma^{\mu}$ and then one obtains the Feynman rule in Eq. (\ref{CT_fermion_split_2}) but without the coefficient $Z_1/Z_3$.

\begin{figure}
\includegraphics[width=1.5cm]{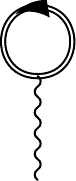}
\caption{One-loop tadpole Feynman diagram.}
\label{Tadpole_1_Loop}
\end{figure}

\begin{figure}
\includegraphics[width=1.5cm]{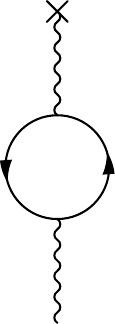}
\caption{First-order, one-loop tadpole Feynman diagram in a background field.}
\label{Tadpole_Pert_no_e-line}
\end{figure}

Now, the renormalization procedure according to Weinberg's Lagrangian density is in this sense very similar to the renormalization procedure in vacuum QED because, the tadpole counterterm corresponding to Eq. (\ref{CT_Tadpole}) with $Z_3=Z_3^{(1)}$ compensates for the divergence of the one-loop tadpole (see Fig. \ref{Tadpole_1_Loop}), which is in agreement with the discussion below Eq. (\ref{A_T_C}).
In fact, by expanding the electron propagator with respect to the background field and by recalling the case of the polarization operator, the only divergent contribution is the one linear in the background field (see Fig. \ref{Tadpole_Pert_no_e-line}). 
This contribution is precisely the one-loop polarization operator with one photon leg represented by the background field and the contribution from Eq. (\ref{CT_Tadpole}) exactly compensates for the resulting divergence (see Fig. \ref{Tadpole_Pert_no_e-line} and also Eq. (\ref{Pi_0})). For higher-order diagrams, the above counterterm systematically removes all the subdivergences due to the one-loop tadpole according to the standard inductive BPHZ iterative method \cite{Itzykson_b_1980}.

\begin{figure}
\includegraphics[width=10cm]{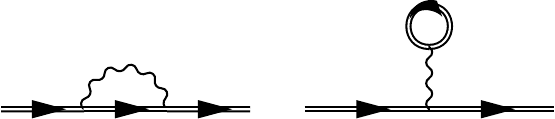}
\caption{Feynman diagrams corresponding to the electron self energy at one loop.}
\label{Fermion_1_loop}
\end{figure}

Let us now go back to the counterterm corresponding to the Feynman rule in Eq. (\ref{CT_fermion_split_2}). We recall that the rule applies to a fermion line because none of the counterterms in the original interaction Lagrangian density in Eq. (\ref{L_i_R}) can renormalize the tadpole diagram itself (this is not a limitation of our approach, where an extra equation is required to renormalize the tadpole by itself, which is Eq. (\ref{A_T_Reno})). Thus, we consider the problem of renormalizing the tadpole when it is attached to a fermion line (note that the resulting diagrams are not 1PI due to the tadpole structure itself) and we start again from the one-loop tadpole diagram in Fig. \ref{Tadpole_1_Loop}, to be thought to be attached to a fermion line. By again imagining to expand it in powers of the background field, we easily conclude that the second counterterm in Eq. (\ref{Counter_SFQED}) at one loop has the right structure and coefficient to remove the corresponding divergence because $(Z^{(1)}_1/Z^{(1)}_3)(Z^{(1)}_3-1)\approx Z^{(1)}_3-1$ at one-loop, as for the Collins' and Weinberg's Lagrangian density. In conclusion, the counterterm $\bar{\psi}(x)\{(Z^{(1)}_2 -1)[i\hat{\partial}-e\hat{\mathcal{A}}(x)]-\delta m^{(1)}+(Z^{(1)}_3-1)e\hat{\mathcal{A}}(x)\}\psi(x)$ corresponding to the Feynman rule in Eqs. (\ref{CT_fermion_1})-(\ref{CT_fermion_split_2}) allows one to renormalize the one-loop electron self energy (see Fig. \ref{Fermion_1_loop}).

In order to understand the Feynman rule in Eq. (\ref{CT_fermion_split_2}) at higher orders, it is first instructive to consider the two-loop case and for this we consider the two-loop electron self energy:
\begin{equation}
\label{eq:two-loop-diagrams}
\begin{aligned}
-i \Sigma^{(2)}_0(x,y \vert \mathcal{A}_0) & = 
\begin{tikzpicture}[baseline={(0,0)}, line width=0.4pt]
    \draw[double, double distance=1pt] (0,0) -- (2.8,0);
    \draw[/tikzfeynman/photon]
        (1.23,0) arc[start angle=180, end angle=0, radius=0.45];
    \draw[/tikzfeynman/photon]
        (0.67,0) arc[start angle=180, end angle=360, radius=0.45];
\end{tikzpicture}
\; + \;
\begin{tikzpicture}[baseline={(0,0)}, line width=0.4pt]
    \draw[double, double distance=1pt] (0,0) -- (2.8,0);
    \draw[/tikzfeynman/photon]
        (0.65,0) arc[start angle=180, end angle=0, radius=0.75];
    \draw[/tikzfeynman/photon]
        (0.95,0) arc[start angle=180, end angle=0, radius=0.45];
\end{tikzpicture}
\; + \;
\begin{tikzpicture}[baseline={(0,0)}, line width=0.4pt]
    \draw[double, double distance=1pt] (0,0) -- (2.8,0);
    \draw[/tikzfeynman/photon]
        (0.70,0) arc[start angle=180, end angle=0, radius=0.70];
    \draw[/tikzfeynman/photon] (1.40,0) -- (1.40,-0.67);
    \draw[double, double distance=1pt]
        (1.40,-0.90) circle[radius=0.23];
\end{tikzpicture}
\\[0.5cm]
&\quad{}+
\begin{tikzpicture}[baseline={(0,0)}, line width=0.4pt]
    \draw[double, double distance=1pt] (0,0) -- (2.8,0);
    \draw[/tikzfeynman/photon]
        (0.40,0) arc[start angle=180, end angle=0, radius=0.40];
    \draw[/tikzfeynman/photon]
        (1.60,0) arc[start angle=180, end angle=0, radius=0.40];
\end{tikzpicture}
\; + \;
2\times\;
\begin{tikzpicture}[baseline={(0,0)}, line width=0.4pt]
    \draw[double, double distance=1pt] (0,0) -- (2.8,0);
    \draw[/tikzfeynman/photon] (0.70,0) -- (0.70,0.67);
    \draw[double, double distance=1pt]
        (0.70,0.90) circle[radius=0.23];
    \draw[/tikzfeynman/photon]
        (1.40,0) arc[start angle=180, end angle=0, radius=0.40];
\end{tikzpicture}
\; + \;
\begin{tikzpicture}[baseline={(0,0)}, line width=0.4pt]
    \draw[double, double distance=1pt] (0,0) -- (2.8,0);
    \draw[/tikzfeynman/photon] (0.95,0) -- (0.95,0.67);
    \draw[double, double distance=1pt]
        (0.95,0.90) circle[radius=0.23];
    \draw[/tikzfeynman/photon] (1.85,0) -- (1.85,0.67);
    \draw[double, double distance=1pt]
        (1.85,0.90) circle[radius=0.23];
\end{tikzpicture}
\\[0.5cm]
&\quad{}+
\begin{tikzpicture}[baseline={(0,0)}, line width=0.4pt]
    \draw[double, double distance=1pt] (0,0) -- (2.8,0);
    \draw[/tikzfeynman/photon]
        (0.45,0) .. controls (0.45,0.50) and (0.75,0.70)
        .. (1.08,0.70);
    \draw[/tikzfeynman/photon]
        (1.72,0.70) .. controls (2.05,0.70) and (2.35,0.50)
        .. (2.35,0);
    \draw[double, double distance=1pt]
        (1.40,0.70) circle[radius=0.32];
\end{tikzpicture}
\; + \;
\begin{tikzpicture}[baseline={(0,0)}, line width=0.4pt]
    \draw[double, double distance=1pt] (0,0) -- (2.8,0);
    \draw[/tikzfeynman/photon] (1.40,0) -- (1.40,0.68);
    \draw[/tikzfeynman/photon] (0.96,1.12) -- (1.84,1.12);
    \draw[double, double distance=1pt]
        (1.40,1.12) circle[radius=0.44];
\end{tikzpicture}
\; + \;
\begin{tikzpicture}[baseline={(0,0)}, line width=0.4pt]
    \draw[double, double distance=1pt] (0,0) -- (2.8,0);
    \draw[/tikzfeynman/photon] (1.40,0) -- (1.40,0.46);
    \draw[/tikzfeynman/photon] (1.40,0.94) -- (1.40,1.41);
    \draw[double, double distance=1pt]
        (1.40,0.70) circle[radius=0.24];
    \draw[double, double distance=1pt]
        (1.40,1.67) circle[radius=0.26];
\end{tikzpicture},
\end{aligned}
\end{equation}      
where for notational simplicity the arrows in the fermion lines are not indicated.

The first step, according to the BPHZ method, is the removal of the one-loop subdivergences. Within Collins' and Weinberg's approach this proceeds as in vacuum because, together with the diagrams already present in vacuum, there is a specific counterterm for the tadpole, which removes the corresponding one-loop divergence in any two-loop diagram. By focusing our attention to the third diagram on the first row and to the third diagram in the last row of Eq. \eqref{eq:two-loop-diagrams}, due to their structure, they require only three subtractions according to the BPHZ method: One involving the tadpole, one involving either the vertex correction (the third diagram in the first row) or the polarization operator (the third diagram in the last row), and one involving their product. The latter contribution for the third diagram in the first row and for the third diagram in the last row read $ie(Z^{(1)}_1-1)(Z^{(1)}_3-1)\hat{\mathcal{A}}(x)$ and $-ie(Z^{(1)}_3-1)^2\hat{\mathcal{A}}(x)$, respectively. These would be the subtractions arising from the BPHZ method applied to the Collins' and Weinberg's Lagrangian densities. However, in our interaction Lagrangian density at one loop, we only have the counterterm $(Z_3^{(1)}-1)\bar{\psi}(x)e\hat{\mathcal{A}}(x)\psi(x)$, which clearly cannot be combined with the counterterms of the vertex correction and of the polarization operator to remove such types of divergences in the fermion line. Therefore, the two-loop counterterm $[(Z^{(1)}_1-1)-(Z^{(1)}_3-1)](Z^{(1)}_3-1)e\bar{\psi}(x)\hat{\mathcal{A}}(x)\psi(x)$ needs to be added ``by hand'' such that the overall counterterm responsible for renormalizing the tadpole divergences at two loops in our interaction Lagrangian density reads
\begin{equation}
\label{CT_mass_op_2}
\begin{split}
&\{(Z^{(1)}_3-1)+[(Z^{(1)}_1-1)-(Z^{(1)}_3-1)](Z^{(1)}_3-1)\}e\bar{\psi}(x)\hat{\mathcal{A}}(x)\psi(x)\\
&\quad\approx \frac{1+(Z^{(1)}_1-1)}{1+(Z^{(1)}_3-1)}(Z^{(1)}_3-1)e\bar{\psi}(x)\hat{\mathcal{A}}(x)\psi(x)=\frac{Z^{(1)}_1}{Z^{(1)}_3}(Z^{(1)}_3-1)e\bar{\psi}(x)\hat{\mathcal{A}}(x)\psi(x).
\end{split}
\end{equation}
This confirms that, in the absence of a counterterm which directly renormalizes the tadpole (in the sense of the one-particle photon correlator), the structure of the counterterm in Eq. (\ref{CT_fermion_split_2}) allows to remove all the divergences induced by the tadpole, when it is attached to a fermion line. In fact, apart from these discussed diagrams, all others can be renormalized similarly in our approach and in Collins' and Weinberg's approach. More explicitly, by applying the Feynman rules in Eqs. (\ref{vertex_CT}), (\ref{photon_CT}), and (\ref{CT_fermion_0}) corresponding to the SFQED Lagrangian density in Eq. (\ref{L_i_R}) to Eq. \eqref{eq:two-loop-diagrams}
one obtains that the quantity $\Sigma^{(2)}_{0}(x,y \vert \mathcal{A}_0)  + \Sigma^{(2)}_{0,\text{c.t.}}(x,y \vert \mathcal{A}_0)$ is finite, where

\begin{equation}
\label{eq:two-loop-counterterms}
\begin{aligned}
-i &\Sigma^{(2)}_{0,\text{c.t.}}(x,y \vert \mathcal{A}_0)  = 
\begin{tikzpicture}[baseline={(0,0)}, line width=0.4pt]
    \draw (0,0.025) -- (2.8,0.025)
          (0,-0.025) -- (2.8,-0.025);
    \draw[fill=white] (1.31,-0.09) rectangle (1.49,0.09);
    \draw (1.31,-0.09) -- (1.49,0.09)
          (1.31,0.09) -- (1.49,-0.09);
\end{tikzpicture}
\\[0.35cm]
&{}+
\begin{tikzpicture}[baseline={(0,0)}, line width=0.4pt]
    \draw (0,0.025) -- (2.8,0.025)
          (0,-0.025) -- (2.8,-0.025);
    \draw[/tikzfeynman/photon]
        (0.80,0) arc[start angle=180, end angle=0, radius=0.60];
    \draw[fill=white] (0.80,0) circle[radius=0.095];
    \draw (0.733,-0.067) -- (0.867,0.067)
          (0.733,0.067) -- (0.867,-0.067);
\end{tikzpicture}
\; + \;
\begin{tikzpicture}[baseline={(0,0)}, line width=0.4pt]
    \draw (0,0.025) -- (2.8,0.025)
          (0,-0.025) -- (2.8,-0.025);
    \draw[/tikzfeynman/photon]
        (0.80,0) arc[start angle=180, end angle=0, radius=0.60];
    \draw[fill=white] (2.00,0) circle[radius=0.095];
    \draw (1.933,-0.067) -- (2.067,0.067)
          (1.933,0.067) -- (2.067,-0.067);
\end{tikzpicture}
\; + \;
\begin{tikzpicture}[baseline={(0,0)}, line width=0.4pt]
    \draw (0,0.025) -- (2.8,0.025)
          (0,-0.025) -- (2.8,-0.025);
    \draw[/tikzfeynman/photon]
        (0.70,0) arc[start angle=180, end angle=0, radius=0.70];
    \draw[fill=white] (1.31,-0.09) rectangle (1.49,0.09);
    \draw (1.31,-0.09) -- (1.49,0.09)
          (1.31,0.09) -- (1.49,-0.09);
\end{tikzpicture}
\; + \;
\begin{tikzpicture}[baseline={(0,0)}, line width=0.4pt]
    \draw (0,0.025) -- (2.8,0.025)
          (0,-0.025) -- (2.8,-0.025);
    \draw[/tikzfeynman/photon] (1.40,0) -- (1.40,-0.62);
    \draw[double, double distance=1pt]
        (1.40,-0.84) circle[radius=0.22];
    \draw[fill=white] (1.40,0) circle[radius=0.095];
    \draw (1.333,-0.067) -- (1.467,0.067)
          (1.333,0.067) -- (1.467,-0.067);
\end{tikzpicture}
\\[0.35cm]
&{}+
\begin{tikzpicture}[baseline={(0,0)}, line width=0.4pt]
    \draw (0,0.025) -- (2.8,0.025)
          (0,-0.025) -- (2.8,-0.025);
    \draw[/tikzfeynman/photon]
        (1.50,0) arc[start angle=180, end angle=0, radius=0.40];
    \draw[fill=white] (0.51,-0.09) rectangle (0.69,0.09);
    \draw (0.51,-0.09) -- (0.69,0.09)
          (0.51,0.09) -- (0.69,-0.09);
\end{tikzpicture}
\; + \;
\begin{tikzpicture}[baseline={(0,0)}, line width=0.4pt]
    \draw (0,0.025) -- (2.8,0.025)
          (0,-0.025) -- (2.8,-0.025);
    \draw[/tikzfeynman/photon]
        (0.50,0) arc[start angle=180, end angle=0, radius=0.40];
    \draw[fill=white] (2.11,-0.09) rectangle (2.29,0.09);
    \draw (2.11,-0.09) -- (2.29,0.09)
          (2.11,0.09) -- (2.29,-0.09);
\end{tikzpicture}
\; + \;
\begin{tikzpicture}[baseline={(0,0)}, line width=0.4pt]
    \draw (0,0.025) -- (2.8,0.025)
          (0,-0.025) -- (2.8,-0.025);
    \draw[fill=white] (0.81,-0.09) rectangle (0.99,0.09);
    \draw (0.81,-0.09) -- (0.99,0.09)
          (0.81,0.09) -- (0.99,-0.09);
    \draw[fill=white] (1.81,-0.09) rectangle (1.99,0.09);
    \draw (1.81,-0.09) -- (1.99,0.09)
          (1.81,0.09) -- (1.99,-0.09);
\end{tikzpicture}
\; + \; 2\times\;
\begin{tikzpicture}[baseline={(0,0)}, line width=0.4pt]
    \draw (0,0.025) -- (2.8,0.025)
          (0,-0.025) -- (2.8,-0.025);
    \draw[/tikzfeynman/photon] (0.75,0) -- (0.75,0.60);
    \draw[double, double distance=1pt]
        (0.75,0.82) circle[radius=0.22];
    \draw[fill=white] (1.91,-0.09) rectangle (2.09,0.09);
    \draw (1.91,-0.09) -- (2.09,0.09)
          (1.91,0.09) -- (2.09,-0.09);
\end{tikzpicture}
\\[0.55cm]
&{}+
\begin{tikzpicture}[baseline={(0,0)}, line width=0.4pt]
    \draw (0,0.025) -- (2.8,0.025)
          (0,-0.025) -- (2.8,-0.025);
    \draw[/tikzfeynman/photon]
        (0.70,0) arc[start angle=180, end angle=0, radius=0.70];
    \draw[fill=white] (1.40,0.70) circle[radius=0.095];
    \draw (1.333,0.633) -- (1.467,0.767)
          (1.333,0.767) -- (1.467,0.633);
\end{tikzpicture}
\; + \;
\begin{tikzpicture}[baseline={(0,0)}, line width=0.4pt]
    \draw (0,0.025) -- (2.8,0.025)
          (0,-0.025) -- (2.8,-0.025);
    \draw[/tikzfeynman/photon] (1.40,0) -- (1.40,0.60);
    \draw[double, double distance=1pt]
        (1.40,0.92) circle[radius=0.32];
    \draw[fill=white] (1.31,1.15) rectangle (1.49,1.33);
    \draw (1.31,1.15) -- (1.49,1.33)
          (1.31,1.33) -- (1.49,1.15);
\end{tikzpicture}
\; + \;
\begin{tikzpicture}[baseline={(0,0)}, line width=0.4pt]
    \draw (0,0.025) -- (2.8,0.025)
          (0,-0.025) -- (2.8,-0.025);
    \draw[/tikzfeynman/photon] (1.40,0) -- (1.40,0.60);
    \draw[double, double distance=1pt]
        (1.40,0.92) circle[radius=0.32];
    \draw[fill=white] (1.40,0.60) circle[radius=0.095];
    \draw (1.333,0.533) -- (1.467,0.667)
          (1.333,0.667) -- (1.467,0.533);
\end{tikzpicture}
\; + \;
\begin{tikzpicture}[baseline={(0,0)}, line width=0.4pt]
    \draw (0,0.025) -- (2.8,0.025)
          (0,-0.025) -- (2.8,-0.025);
    \draw[/tikzfeynman/photon] (1.40,0) -- (1.40,0.76);
    \draw[double, double distance=1pt]
        (1.40,1.00) circle[radius=0.24];
    \draw[fill=white] (1.40,0.40) circle[radius=0.095];
    \draw (1.333,0.333) -- (1.467,0.467)
          (1.333,0.467) -- (1.467,0.333);
\end{tikzpicture},
\end{aligned}
\end{equation}      
with the first counterterm being the two-loop one and all the others being at one loop. The counterterm corresponding to the contribution in Eq. (\ref{CT_mass_op_2}) is contained in the first diagram.

The two-loop result above can be extended to higher orders but the calculations would become cumbersome and we refer to the discussion below Eq. (\ref{Z_C_Z}), which ensures already the complete equivalence between the renormalization procedure based on Collins' and Weinberg's Lagrangian densities and on our Lagrangian density. 

Finally, the problem of the renormalization of the tadpole can also be investigated within the so-called self-consistent formulation of SFQED, where the fermion line is dressed not with the background field but with the total vacuum electromagnetic field \cite{Brouder_2002}. Looking at Eq. (\ref{G_00T}), one can introduce for this propagator the Feynman rule
\begin{equation}
\label{Tot_prop}
\begin{tikzpicture}[baseline={([yshift=-0.5ex]current bounding box.center)}]
    \draw[double, double distance=1pt, line width=1.2pt] (1.8,0) -- (0,0);
    \draw[double, double distance=1pt, line width=1.2pt, arrows = {-Latex[width=10pt, length=10pt]}] (2.8,0) -- (1.2,0);
\end{tikzpicture}=iG^{(0)}_{T,0}(x,y|\mathcal{A}_{T,0}),
\end{equation}
with the double thick line. This notation is analogous to the one introduced for the dressed propagator as a function of the background field $\mathcal{A}_0^\mu(x)$ (see Fig. \ref{Dressed_Prop}). Indeed, $G^{(0)}_{T,0}(x,y|\mathcal{A}_{T,0})$ is understood as an infinite series of tree-level insertions of the total vacuum electromagnetic field $\mathcal{A}_{T,0}^\mu(x)$ (see also Fig. \ref{Tadpole_Exact})
\begin{equation}
\label{FD_Dressed_Prop}
\begin{tikzpicture}[baseline={([yshift=-0.5ex]current bounding box.center)}]
    \draw[double, double distance=1pt, line width=1.2pt] (1.8,0) -- (0,0);
    \draw[double, double distance=1pt, line width=1.2pt, arrows = {-Latex[width=10pt, length=10pt]}] (2.8,0) -- (1.2,0);
\end{tikzpicture}
\;=\;
\begin{tikzpicture}[baseline=-0.5ex]
\begin{feynman}
\vertex (b1) at (0,0);
\vertex (b2) at (2.4,0);
\diagram*{
(b1) -- [anti fermion] (b2)
};
\end{feynman}
\end{tikzpicture}
\;+\;
\begin{tikzpicture}[baseline=-0.5ex]
\begin{feynman}
\vertex (c1) at (0,0);
\vertex (c2) at (1.2,0);
\vertex (c3) at (2.4,0);
\vertex (cx) at (1.2,1);
\diagram*{
(c1) -- [anti fermion] (c2)
     -- [anti fermion] (c3),
(c2) -- [photon,very thick] (cx)
};
\end{feynman}
\draw[very thick]
([xshift=-2.5pt,yshift=-2.5pt]cx)
--
([xshift=2.5pt,yshift=2.5pt]cx);
\draw[very thick]
([xshift=-2.5pt,yshift=2.5pt]cx)
--
([xshift=2.5pt,yshift=-2.5pt]cx);
\end{tikzpicture}
\;+\;
\begin{tikzpicture}[baseline=-0.5ex]
\begin{feynman}
\vertex (d1) at (0,0);
\vertex (d2) at (0.8,0);
\vertex (d3) at (1.6,0);
\vertex (d4) at (2.4,0);
\vertex (dx1) at (0.8,1);
\vertex (dx2) at (1.6,1);
\diagram*{
(d1) -- [anti fermion](d2)
     -- [anti fermion] (d3)
     -- [anti fermion] (d4),
(d2) -- [photon,very thick] (dx1),
(d3) -- [photon,very thick] (dx2)
};
\end{feynman}
\draw[very thick]
([xshift=-2.5pt,yshift=-2.5pt]dx1)
--
([xshift=2.5pt,yshift=2.5pt]dx1);
\draw[very thick]
([xshift=-2.5pt,yshift=2.5pt]dx1)
--
([xshift=2.5pt,yshift=-2.5pt]dx1);
\draw[very thick]
([xshift=-2.5pt,yshift=-2.5pt]dx2)
--
([xshift=2.5pt,yshift=2.5pt]dx2);
\draw[very thick]
([xshift=-2.5pt,yshift=2.5pt]dx2)
--
([xshift=2.5pt,yshift=-2.5pt]dx2);
\end{tikzpicture}
\;+\;\cdots,
\end{equation}
where the thick photon lines stand for total vacuum electromagnetic field insertions:
\begin{equation}
\begin{tikzpicture}[baseline={([yshift=-0.5ex]current bounding box.center)}]
    \begin{feynman}
        \vertex (a1);
        \vertex [right=1cm of a1] (a2);
        \diagram*{
            (a1) -- [photon, very thick] (a2),
        };
        \node[inner sep=0pt] at (a1) {$\boldsymbol{\times}$};
    \end{feynman}
\end{tikzpicture}
\;=\;
\begin{tikzpicture}[baseline={([yshift=-0.5ex]current bounding box.center)}]
    \begin{feynman}
        \vertex (a1);
        \vertex [right=1cm of a1] (a2);
        \diagram*{
            (a1) -- [photon] (a2),
        };
        \node[inner sep=0pt] at (a1) {$\times$};
    \end{feynman}
\end{tikzpicture}
\;+\; 
\begin{tikzpicture}[baseline={([yshift=-0.5ex]current bounding box.center)}]
    \filldraw[fill=gray, draw=black, line width=0.4pt]
        (1.35,0) circle[radius=0.4];
    \draw[/tikzfeynman/photon] (1.75,0) -- (2.7,0);
\end{tikzpicture}\-\
.
\end{equation}

The self-consistent formulation of SFQED is convenient because all the tadpoles are absorbed in the definition of the total vacuum electromagnetic field such that the diagrammatic expansions of the correlators have the same topological structure as those in vacuum QED. This implies that the renormalization within the self-consistent formulation of SFQED is carried out exactly as in vacuum QED, and only the total vacuum electromagnetic field (and then the tadpole) is renormalized separately.

\section{Additional remarks on the tadpole in the presence of a free background field}
\label{Remarks_Tadpole}
The main novelty of SFQED in the renormalization analysis carried out above is the presence of the tadpole and its relation to the renormalization of the background field. For this reason, we would like to make some additional remarks about this quantity related in particular to the case of free background fields such as a constant field or a plane-wave field.  Although a free background field with no sources anywhere is an idealization, these remarks are useful because free fields like constant fields or plane-wave field are widely employed as insightful models of more realistic fields.

We go back to Eq. (\ref{A_T-R-partV}) and we use the observation below that equation on the transverse structure of the polarization operator and of the $n$-point photon correlation functions with $n\ge 2$ to make the replacement
\begin{equation}
D_0^{(0),\mu\nu}(x-y|0)\to\Delta_0^{(0)}(x-y|0)\eta^{\mu\nu},
\end{equation}
where
\begin{equation}
\Delta_0^{(0)}(x-y|0)=\int\frac{d^4k}{(2\pi)^4}\frac{e^{-i(k(x-y))}}{k^2+i0}
\end{equation}
is the free, tree-level scalar propagator. Also, we can replace all the derivatives with respect to $y^{\mu}$ into derivatives with respect to $x_1^{\mu}$ in the first two lines of that equation and use the fact that the field $\mathcal{A}_{T,0}^{\mu}(x)$ satisfies the Lorenz-gauge condition to obtain
\begin{equation}
\label{A_T_free}
\begin{split}
&\mathcal{A}^{\mu}_{T,0}(x)=\mathcal{A}_0^{\mu}(x)-\frac{1}{Z_3}\int d^4 x_1 d^4 y\Delta\Pi_T(y-x_1|0)\Delta_0^{(0)}(x-y|0)\square_{x_1}\mathcal{A}_{T,0,\mu}(x_1)\\
&\quad-\frac{Z_3 - 1}{Z_3}\int d^4 x_1 d^4 y\,\delta^4(y-x_1)\Delta_0^{(0)}(x-y|0)\square_{x_1}\mathcal{A}_{T,0}^{\mu}(x_1)\\
&\quad+\sum_{n=2}^{\infty}\frac{1}{n!}\int d^4x_1\cdots d^4x_n d^4y\,\Delta_0^{(0)}(x-y|0)\Pi^{\mu}_{T,0,\nu_1\cdots\nu_n}(y,x_1,\ldots,x_n|0)\mathcal{A}_{T,0}^{\nu_1}(x_1)\cdots \mathcal{A}_{T,0}^{\nu_n}(x_n),
\end{split}
\end{equation}
where we purposely wrote the free scalar propagator next to the wave operator acting on the total vacuum electromagnetic field. In fact, we observe that this equation is ambiguous in the case of the total vacuum electromagnetic field satisfying the free Maxwell's equations (if this is not the case, we recover the result in Eq. (\ref{A_T_R_2})), which can also be seen transparently passing to momentum space. However, the ambiguity is resolved by noticing that if we would first let the wave operator $\square_{x_1}$ act on the total vacuum electromagnetic field $\mathcal{A}_{T,0}^{\mu}(x_1)$, the second and the third terms in Eq. (\ref{A_T_free}) would vanish and the resulting equation would not be renormalizable. Thus, our prescription in the case of a background electromagnetic field leading to a free total vacuum electromagnetic field is to assume that the latter is actually not free and to take the free-field limit only at the end of the calculations. In Eq. (\ref{A_T_free}) this ultimately amounts to perform the substitution $\Delta_0^{(0)}(x-y|0)\square_{x_1}\to -\delta^4(x-y)$. 

Determining a priori whether the total vacuum electromagnetic field is exactly free or not may clearly be impossible. This occurs, however, in the two theoretically important cases of a background constant field and of a background plane-wave field, which are free fields themselves. Both in a constant field and in a plane-wave field, in fact, one can conclude that the renormalized vacuum four-current density has to vanish based on general considerations: The vacuum four-current density is a gauge-invariant four-vector and in either a constant field and or in a plane-wave field one cannot construct a gauge-invariant four-vector only out of the fields themselves (in a plane wave one can also use the wave four-vector, which is proportional to the null quantity $n^{\mu}=(1,\bm{n})$, with $\bm{n}$ being the direction of propagation of the plane wave but, since $n^{\mu}$ is perpendicular to the plane-wave tensor field, the above conclusion still holds). This result has been shown explicitly both in a constant field and in a plane-wave field at one loop in Ref. \cite{Ahmadiniaz_2019} (for the plane-wave case, see also the appendix).

It is instructive to see explicitly at one loop how the above-mentioned ambiguity arises and how the requirement of renormalizability allows one to solve the ambiguity itself. Since the divergence in the tadpole arises in the term linear in the background field (see Fig. \ref{Tadpole_Pert_no_e-line}), it is sufficient here to calculate the one-loop total vacuum electromagnetic field at the linear order with respect to the background field. Such a perturbative calculation can be carried out for an arbitrary background field $\mathcal{A}^{\mu}(x)$ and therefore we first assume that, although $\mathcal{A}^{\mu}(x)$ satisfies the Lorenz gauge condition, it does not necessarily satisfy the free Maxwell's equations. 

We start from Eq. (\ref{A_phys}), which at one loop reads
\begin{equation}
\mathcal{A}^{(1),\mu}_{T,0}(x)=\mathcal{A}_0^{\mu}(x)+ie_0\int d^4yD_0^{(0),\mu\nu}(x-y|0)\text{Tr}(\gamma_{\nu}G^{(0)}_0(y,y|\mathcal{A}_0)).
\end{equation}
Now, we expand this expression up to linear terms in the background field. By using Eq. (\ref{trace derivative Gtot}), we obtain
\begin{equation}
\mathcal{A}^{(1),\mu}_{T,0}(x)=\mathcal{A}_0^{\mu}(x)+\int d^4yd^4zD_0^{(0),\mu\nu}(x-y|0)\Pi^{(1)}_{T,0,\nu\lambda}(y-z|0)\mathcal{A}_0^{\lambda}(z),
\end{equation}
where we have already exploited the fact that the tadpole in vacuum vanishes. At this point, we pass to the renormalized quantities, we use again the transversality of the polarization operator and the fact that the background field satisfies the Lorenz-gauge condition, and we obtain (see Eq. (\ref{Pi_0}))
\begin{equation}
\label{A_phys_1}
\begin{split}
\mathcal{A}^{(1),\mu}_T(x)&=\mathcal{A}^{\mu}(x)(1+\delta_3^{(1)})-\int d^4yd^4z\Delta_0^{(0)}(x-y|0)\square_y\Delta \Pi^{(1)}_T(y-z|0)\mathcal{A}^{\lambda}(z)\\
&\quad+\delta_3^{(1)}\int d^4yd^4z\Delta_0^{(0)}(x-y|0)\square_y\delta^4(y-z)\mathcal{A}^{\lambda}(z),
\end{split}
\end{equation}
where $\delta_3^{(1)}=1-Z_3^{(1)}$.

If we now integrate by parts twice in $d^4y$, we again see that the infinite quantities cancel out and, by using the expression of the one-loop vacuum polarization operator in momentum space \cite{Schwartz_b_2014}, we obtain the final expression of the one-loop total vacuum electromagnetic field in the form
\begin{equation}
\label{A_T_L}
\mathcal{A}_T^{(1),\mu}(x)=\mathcal{A}^{\mu}(x)+\frac{2\alpha}{\pi}\int\frac{d^4q}{(2\pi)^4}e^{-i(qx)}\check{\mathcal{A}}^{\mu}(q)\int_0^1 du \,u(1-u)\log\left(\frac{m^2-u(1-u)q^2-i0}{m^2}\right),
\end{equation}
where $\check{\mathcal{A}}^{\mu}(q)$ is the Fourier transform of the background field. This equation shows that for a free background electromagnetic field ($\check{\mathcal{A}}^{\mu}(q)\propto \delta(q^2)$) the correction linear in the background field vanishes. Concerning higher orders terms in the background field, it has been shown that they also vanish in a plane wave \cite{Ahmadiniaz_2019,Di_Piazza_2022_b}, whereas they do not, for example, in a constant magnetic field \cite{Ahmadiniaz_2019}. In the case of a plane wave, one can obtain an even more general result. In fact, by computing the exact total vacuum background tensor field $\mathcal{F}_T^{\mu\nu}(x)=\partial^{\mu}\mathcal{A}_T^{\nu}(x)-\partial^{\nu}\mathcal{A}_T^{\mu}(x)$, one can again show that due to gauge invariance and to the fact that both the field invariants vanish for a plane wave, $\mathcal{F}_T^{\mu\nu}(x)$ is equal to the field tensor of the plane wave itself. Instead, a prominent case of a non-free field is that of the Coulomb field for which the zero component of the quantum correction is different from zero and it can be shown that the second term in Eq. (\ref{A_T_L}) reduces to the well-known expression of the Uehling potential \cite{Uehling_1935,Greiner_b_2009}.

Now, going back to the case of a free background field, we also see in Eq. (\ref{A_phys_1}) that we can write $\square_y=\square_z$ and integrate by parts in $d^4z$. The conclusion would again be that the equation could not be renormalized. Thus, in the case of a free background field the prescription is again to first consider the field as being generated by a non-zero background four-current density and only after the renormalization is carried out, take the limit of a free field. In the case of Eq. (\ref{A_phys_1}) this amounts to perform the substitution $\Delta_0^{(0)}(x-y|0)\square_y\to-\delta^4(x-y)$. 

Finally, we note that the above prescription also solves a possible inconsistency in Collins' Lagrangian density in Eq. (\ref{L_B_C}), which would not feature a counterterm for the tadpole if $\mathcal{J}^{\mu}(x)=0$. Interestingly,  Weinberg's Lagrangian density does not feature this shortcoming but only because, as compared to Collins' Lagrangian density, the counterterm responsible for the renormalization of the tadpole is written in terms of the background electromagnetic field (see Eq. (\ref{L_B_W})). These two counterterms in the Collins' and Weinberg's Lagrangian densities, however, differ only by a total four-derivative such that, as already discussed, the two Lagrangian densities are equivalent. This is another indication that such kinds of ambiguities occur in the idealized case of a free background field as the latter does not vanish despite the fact that the background four-current density is identically zero.

\section{Conclusion}\label{sec-conclusion}

In this work, we have studied the renormalization of QED in background fields from the standpoint of the so-called renormalized perturbation theory, i.e., by introducing the renormalization constants $\delta m$, $Z_1$, $Z_2$, and $Z_3$ and the relative counterterms. As a result of a perturbative expansion of the total vacuum electromagnetic field $\A^\mu_{T,0}(x)$, we have shown that, if one uses the unrenormalized Lagrangian density in Eq. (\ref{Lb}), the renormalization condition $\mathcal{A}^{\mu}(x)=\sqrt{Z_3}\mathcal{A}_0^{\mu}(x)$ emerges for the background field, in agreement with Refs. \cite{Braun_1972,Braun_1978}. 

To the best of our knowledge, two approaches have been proposed to the renormalization of SFQED: the one followed in this work, where the background field is introduced already in the bare Lagrangian density \cite{Braun_1978,Brouder_2002} and the one where the external field is introduced in the renormalized Lagrangian density \cite{Collins_b_1984,Weinberg_b_1_2005}. We have shown explicitly that these two approaches are equivalent after shifting the quantum radiation field by a counterterm proportional to the background electromagnetic field. Even though in the second approach the unrenormalized background field is never explicitly mentioned, the way how the renormalized background field is introduced implies that it is actually renormalized as the electromagnetic field rather than as a charge. We have shown that there is no contradiction with our renormalization procedure because the starting unrenormalized Lagrangian density would be different than that in Eq. (\ref{Lb}). In particular, while in our approach the unrenormalized background field and four-current density fulfill Maxwell's equations, this is not the case in the approach of Refs. \cite{Collins_b_1984,Weinberg_b_1_2005}. This is however not a significant difference from a physical point of view because in both approaches the renormalized background field and four-current density do fulfill Maxwell's equations.

Furthermore, we have derived the Feynman rules for the counterterms in SFQED with a particular emphasis on the new counterterm as compared to vacuum QED, which is necessary to renormalize the tadpoles. In this respect, we have shown that the case of a free background field needs to be treated with care in relation to the renormalization procedure. Due to possible ambiguities arising in such idealized background fields, we have put forward the prescription to assume that the background four-current density does not vanish until the renormalization procedure is carried out.

\begin{acknowledgments}
A.D.P. is partially supported by the U.S. National Science Foundation Mid-scale Research Infrastructure Program under Award No. PHY-2329970.

This material is based upon work supported by the U.S. Department of Energy [National Nuclear Security Administration] University of Rochester ``National Inertial Confinement Fusion Program'' under Award Number DE-NA0004144.

This report was prepared as an account of work sponsored by an agency of the United States Government. Neither the United States Government nor any agency thereof, nor any of their employees, makes any warranty, express or implied, or assumes any legal liability or responsibility for the accuracy, completeness, or usefulness of any information, apparatus, product, or process disclosed, or represents that its use would not infringe privately owned rights. Reference herein to any specific commercial product, process, or service by trade name, trademark, manufacturer, or otherwise does not necessarily constitute or imply its endorsement, recommendation, or favoring by the United States Government or any agency thereof. The views and opinions of authors expressed herein do not necessarily state or reflect those of the United States Government or any agency thereof.

We acknowledge useful discussions with A. Hosak and we thank T. de Vos for his valuable comments. A.D.P. also acknowledges insightful discussions with C. Brouder, V. M. Shabaev, S. Volkov, and V. A. Yerokhin.

\end{acknowledgments}

%
\appendix

\section{Re-evaluation of the one-loop vacuum four-current density in an arbitrary plane wave}\label{sec-tadpole-pw}

In Ref. \cite{Di_Piazza_2022_b}, the one-loop vacuum four-current density in an arbitrary plane wave was computed and it was finally expressed as a divergent four-dimensional integral over the four-momentum flowing in the loop, which however features some ambiguities (see Fig. \ref{Tadpole_1_Loop}, where the double plain line indicates the Volkov propagator, and notice that in Ref. \cite{Di_Piazza_2022_b} the vacuum four-current density was defined with the opposite sign as here).

In this appendix, we would like to show that these ambiguities can actually be resolved and we conclude that the vacuum four-current density actually vanishes in agreement with Ref. \cite{Ahmadiniaz_2019}. Although this has no physical implications because the ultimate conclusion is still that the quantum correction to the background electromagnetic field, i.e., the tadpole, vanishes in a plane wave after renormalization, we think that it is instructive to re-derive the one-loop vacuum four-current density in an arbitrary plane wave. In order to simplify the comparison with Ref. \cite{Di_Piazza_2022_b}, we work with unrenormalized quantities.

We recall that the vacuum four-current density $\mathcal{J}_{v,0}^{\mu}(x)$ is defined as $iT_0^{\mu}(x)$, where $T_0^{\mu}(x)$ is the tadpole amplitude in Eq. (\ref{T_0}). In the original paper \cite{Schwinger_1951}, Schwinger only specifies that starting from the propagator $G_0(x,y|\mathcal{A}_0)$ the limit $y\to x$ has to be performed by taking the average of the forms obtained by letting $y$ approach $x$ ``from the future and from the past''. Since the limit under consideration is four-dimensional, a more precise prescription should be provided. From the wording used by Schwinger, one might conclude that one first should take the limit on the space coordinates $\bm{y}\to\bm{x}$ and then the remaining symmetric limit $(\lim_{t_y\to t_x^+}+\lim_{t_y\to t_x^-})/2$ of the resulting expression \cite{Di_Piazza_2022_b}. On the other hand, in Refs. \cite{Greiner_b_1985,Brouder_2002} the complementary prescription is proposed, where one first takes the coincidence limit in the time coordinate (in the average sense specified above) and then in the space coordinates (in the latter case without a further prescription). However, in the later publication \cite{Schwinger_1959}, Schwinger notices that ``for the purpose of evaluating the commutator [of the Dirac four-current density] with the charge density, it suffices to consider the operator product at distinct spatial points and equal times''. 

The problem of taking the coincidence limit in the vacuum four-current density is similar to the corresponding problem to compute the vacuum energy-momentum tensor in quantum field theory in curved spacetimes \cite{Birrell_b_1982}. Within the so-called point-splitting method \cite{Christensen_1976} applied to our case, one would first define
\begin{equation}
T_0^{(1),\mu}(x;\epsilon)=-\frac{e_0}{2}\text{Tr}((\gamma^{\mu}[U(x,x+\epsilon)G^{(0)}_0(x+\epsilon,x|\mathcal{A}_0)+U(x+\epsilon,x)G^{(0)}_0(x,x+\epsilon|\mathcal{A}_0)])),
\end{equation}
where $\epsilon^{\mu}$ is a constant, infinitesimal four-vector and where the Wilson line
\begin{equation}
U(x,y)=e^{-ie_0\int_y^xdz_{\mu}\mathcal{A}_0^{\mu}(z)}
\end{equation}
is included in each term to preserve gauge invariance in the limiting procedure \cite{Boulware_1966}. Indeed, at the end of the calculation one has to take the limit $\epsilon^{\mu}\to 0$ and in a covariant way, i.e., in a way to keep the manifest covariance of the equations. This can be quite complicated in the case of the energy-momentum tensor in quantum field theory in curved spacetimes but in the present case we will show that the general structure of this quantity for an infinitesimal $\epsilon^{\mu}$ is $T_0^{(1),\mu}(x;\epsilon)\approx \mathcal{T}_{0,\lambda\rho}^{(1),\mu}(x;0)\epsilon^{\lambda}\epsilon^{\rho}/\epsilon^2$ plus terms vanishing for $\epsilon^{\mu}\to 0$ (following an arbitrary prescription here). At this point the covariant ``limit'' is taken by replacing $\epsilon^{\mu}\epsilon^{\nu}/\epsilon^2\to \eta^{\mu\nu}/4$ \cite{Peskin_b_1995}. Below, we will show that this procedure, concisely indicated as $T_0^{(1),\mu}(x)=\lim_{\epsilon^{\mu}\to 0}T_0^{(1),\mu}(x;\epsilon)$, leads to the result $T_0^{(1),\mu}(x)=0$. We stress the fact that the limit has to be taken in the propagator expressed in coordinates space.

In order to start the evaluation of the one-loop tadpole in an arbitrary plane wave, we consider the expression of the Volkov propagator $G_{V,0}(x,y)$ in coordinates space found in Refs. \cite{Brown_1964,Di_Piazza_2018_d}:
\begin{equation}
\begin{split}
G^{(0)}_{V,0}(x,y)=&-\frac{i}{16\pi^2}\left[i\gamma^{\mu}\frac{\partial}{\partial x^{\mu}}-e_0\hat{\mathcal{A}}_0(\phi_x)+m_0\right]e^{-ie_0(x-y)^{\nu}\int_0^1du\, \mathcal{A}_{0,\nu}(\phi_y+u(\phi_x-\phi_y))}\\
&\times\int_0^{\infty}\frac{ds}{s^2}\,\bigg[1+e_0s\hat{n}\frac{\hat{\mathcal{A}}_0(\phi_x)-\hat{\mathcal{A}}_0(\phi_y)}{\phi_x-\phi_y}\bigg]e^{-is\tilde{m}_0^2(\phi_x,\phi_y)-i(x-y)^2/4s}.
\end{split}
\end{equation}
In this expression, the light-cone time coordinates $\phi_x=(nx)$ and $\phi_y=(ny)$ have been introduced, with $n^{\mu}=(1,\bm{n})$ being a four-dimensional quantity characterizing the plane-wave propagation direction such that $n^2=0$. The unrenormalized plane-wave field is indicated as $\mathcal{A}_0^{\mu}(\phi_x)$ and, assuming to work within the Lorenz gauge with the initial condition $\lim_{\phi_x\to-\infty}\mathcal{A}_0^{\mu}(\phi_x)=0$, it is such that $(n\mathcal{A}_0(\phi_x))=0$. Finally, the so-called squared dressed mass
\begin{equation}
\label{m_tilde}
\tilde{m}_0^2(\phi_x,\phi_y)=m_0^2-e_0^2\int_0^1du\, \mathcal{A}_0^2(\phi_y+u(\phi_x-\phi_y))+e_0^2\left[\int_0^1du\, \mathcal{A}_0^{\mu}(\phi_y+u(\phi_x-\phi_y))\right]^2
\end{equation}
has also been introduced.

We first compute the trace $K_0^{\mu}(x,y|\mathcal{A}_0)=\text{Tr}(\gamma^{\mu}U(y,x)G^{(0)}_{V,0}(x,y))$ noticing that in a plane wave it is
\begin{equation}
U(x,y)=e^{-ie_0(x-y)^{\nu}\int_0^1du\, \mathcal{A}_{0,\nu}(\phi_y+u(\phi_x-\phi_y))}\,,
\end{equation}
and we already see that the term in $G_{V,0}(x,y)$ proportional to $m_0$ does not contribute:
\begin{equation}
\begin{split}
K_0^{\mu}(x,y|\mathcal{A}_0)&=-\frac{i}{4\pi^2}U(y,x)\left[i\frac{\partial}{\partial x^{\lambda}}-e_0\mathcal{A}_{0,\lambda}(\phi_x)\right]U(x,y)\int_0^{\infty}\frac{ds}{s^2}\,e^{-is\tilde{m}_0^2(\phi_x,\phi_y)-i(x-y)^2/4s}\\
&\qquad\times\bigg\{\eta^{\mu\lambda}+e_0s\frac{-n^{\mu}[\mathcal{A}^{\lambda}_0(\phi_x)-\mathcal{A}^{\lambda}_0(\phi_y)]+n^{\lambda}[\mathcal{A}^{\mu}_0(\phi_x)-\mathcal{A}^{\mu}_0(\phi_y)]}{\phi_x-\phi_y}\bigg\}\\
&=\frac{1}{4\pi^2}U(y,x)\frac{\partial}{\partial x^{\lambda}}U(x,y)\int_0^{\infty}\frac{ds}{s^2}\,e^{-is\tilde{m}_0^2(\phi_x,\phi_y)-i(x-y)^2/4s}\\
&\qquad\times\bigg\{\eta^{\mu\lambda}+e_0s\frac{-n^{\mu}[\mathcal{A}^{\lambda}_0(\phi_x)-\mathcal{A}^{\lambda}_0(\phi_y)]+n^{\lambda}[\mathcal{A}^{\mu}_0(\phi_x)-\mathcal{A}^{\mu}_0(\phi_y)]}{\phi_x-\phi_y}\bigg\}\\
&\quad+\frac{ie_0}{4\pi^2}\int_0^{\infty}\frac{ds}{s^2}\,e^{-is\tilde{m}_0^2(\phi_x,\phi_y)-i(x-y)^2/4s}\\
&\qquad\times\bigg\{\mathcal{A}_0^{\mu}(\phi_x)-e_0s\frac{n^{\mu}(\mathcal{A}_0(\phi_x)[\mathcal{A}_0(\phi_x)-\mathcal{A}_0(\phi_y)])}{\phi_x-\phi_y}\bigg\},
\end{split}
\end{equation}
where the partial derivative $\partial/\partial x^{\lambda}$ acts on everything on its right.

Since the coincidence limit has to be performed at the end of the calculation, we now compute the derivatives in $G^{(0)}_{V,0}(x,y)$. After noticing that the partial derivative $\partial/\partial x^{\lambda}$ of the function in the braces vanishes, we only need two quantities:
\begin{align}
\begin{split}
\mathcal{D}_{1,\lambda}(x,y|\mathcal{A}_0)&=U(y,x)\frac{\partial}{\partial x^{\lambda}}U(x,y)\\
&=e^{ie_0(x-y)^{\nu}\int_0^1du\, \mathcal{A}_{0,\nu}(\phi_y+u(\phi_x-\phi_y))}\frac{\partial}{\partial x^{\lambda}}e^{-ie_0(x-y)^{\nu}\int_0^1du\, \mathcal{A}_{0,\nu}(\phi_y+u(\phi_x-\phi_y))}\\
&=-ie_0\int_0^1du \left[\mathcal{A}_{0,\lambda}(\phi_y+u(\phi_x-\phi_y))+n_{\lambda}u ((x-y)\mathcal{A}'_0(\phi_y+u(\phi_x-\phi_y)))\right],
\end{split}\\
\begin{split}
\mathcal{D}_{2,\lambda}(s,x,y|\mathcal{A}_0)&=e^{is\tilde{m}_0^2(\phi_x,\phi_y)+i(x-y)^2/4s}\frac{\partial}{\partial x^{\lambda}}e^{-is\tilde{m}_0^2(\phi_x,\phi_y)-i(x-y)^2/4s}\\
&=-i\left[sn_{\lambda}\frac{\partial\tilde{m}_0^2(\phi_x,\phi_y)}{\partial \phi_x}+\frac{x_{\lambda}-y_{\lambda}}{2s}\right],
\end{split}
\end{align}
where the prime indicates the derivative with respect to the argument and 
\begin{equation}
\label{D_m2}
\begin{split}
\frac{\partial\tilde{m}_0^2(\phi_x,\phi_y)}{\partial \phi_x}&=2e_0^2\left[-\int_0^1du\,u (\mathcal{A}_0(\phi_y+u(\phi_x-\phi_y))\mathcal{A}'_0(\phi_y+u(\phi_x-\phi_y)))\right.\\
&\quad\left.+\int_0^1du\, \mathcal{A}_0^{\mu}(\phi_y+u(\phi_x-\phi_y))\int_0^1du\,u \mathcal{A}'_{0,\mu}(\phi_y+u(\phi_x-\phi_y))\right].
\end{split}
\end{equation}
The final expression of the quantity $K_0^{\mu}(x,y|\mathcal{A}_0)$ reads
\begin{equation}
\label{T_xy}
\begin{split}
&K_0^{\mu}(x,y|\mathcal{A}_0)=\frac{1}{4\pi^2}\int_0^{\infty}\frac{ds}{s^2}\,e^{-is\tilde{m}_0^2(\phi_x,\phi_y)-i(x-y)^2/4s}\\
&\qquad\times\bigg\{\mathcal{D}_1^{\mu}(x,y|\mathcal{A}_0)-e_0sn^{\mu}\frac{(\mathcal{D}_1(x,y|\mathcal{A}_0)[\mathcal{A}_0(\phi_x)-\mathcal{A}_0(\phi_y)])}{\phi_x-\phi_y}\bigg\}\\
&\quad+\frac{1}{4\pi^2}\int_0^{\infty}\frac{ds}{s^2}\,e^{-is\tilde{m}_0^2(\phi_x,\phi_y)-i(x-y)^2/4s}\\
&\qquad\times\bigg\{\mathcal{D}_2^{\mu}(s,x,y|\mathcal{A}_0)-i\frac{e_0}{2}[\mathcal{A}^{\mu}_0(\phi_x)-\mathcal{A}^{\mu}_0(\phi_y)]+i\frac{e_0}{2}n^{\mu}\frac{((x-y)[\mathcal{A}_0(\phi_x)-\mathcal{A}_0(\phi_y))]}{\phi_x-\phi_y}\bigg\}\\
&\quad+\frac{ie_0}{4\pi^2}\int_0^{\infty}\frac{ds}{s^2}\,e^{-is\tilde{m}_0^2(\phi_x,\phi_y)-i(x-y)^2/4s}\\
&\qquad\times\bigg\{\mathcal{A}_0^{\mu}(\phi_x)-e_0s\frac{n^{\mu}(\mathcal{A}_0(\phi_x)[\mathcal{A}_0(\phi_x)-\mathcal{A}_0(\phi_y)])}{\phi_x-\phi_y}\bigg\}.
\end{split}
\end{equation}
At this point, we need to take the limit 
\begin{equation}
\lim_{\epsilon^{\mu}\to 0}[K_0^{\mu}(x+\epsilon,x|\mathcal{A}_0)+K_0^{\mu}(x,x+\epsilon|\mathcal{A}_0)]
\end{equation}
in the sense indicated above. However, we can first make a substantial simplification in the quantities $K_0^{\mu}(x+\epsilon,x|\mathcal{A}_0)$ and $K_0^{\mu}(x,x+\epsilon|\mathcal{A}_0)$ by observing that the terms featuring a pre-exponential dependence on $s$ as $1/s$ times a function that vanishes in the coincidence limit will ultimately vanish. The reason is that in the limit of small values of $\epsilon^{\mu}$ the integral in $s$ in this case will be proportional to either the Hankel function $\text{H}^{(2)}_0(m\sqrt{\epsilon^2})$ if $\epsilon^2>0$ or to the modified Bessel function $\text{K}_0(m\sqrt{-\epsilon^2})$ if $\epsilon^2<0$ \cite{Gradshteyn_b_2000}. Since both these functions diverge logarithmically in the limit of vanishing argument, the contributions of the mentioned terms vanish (rigorously speaking, assuming that the pre-exponential function under discussion vanishes faster than $1/\log(|\epsilon^2|)$, which is always the case). Moreover, Eq. (\ref{D_m2}) shows that the partial derivative $\partial\tilde{m}_0^2(\phi_x,\phi_y)/\partial \phi_x$ vanishes in the coincidence limit such that the first term in $\mathcal{D}_2^{\mu}(s,x,y|\mathcal{A}_0)$ does not contribute. Finally, by combining the second term in the second line and the second term in the last line, one can also show that their sum vanishes in the coincidence limit. Therefore, with the purpose of computing the coincidence limit, we can equivalently analyze the quantity
\begin{equation}
\begin{split}
&\tilde{K}_0^{\mu}(x,y|\mathcal{A}_0)=\frac{1}{4\pi^2}\int_0^{\infty}\frac{ds}{s^2}\,e^{-is\tilde{m}_0^2(\phi_x,\phi_y)-i(x-y)^2/4s}\\
&\quad\times\bigg\{\mathcal{D}_1^{\mu}(x,y|\mathcal{A}_0)-i\frac{x^{\mu}-y^{\mu}}{2s}+i\frac{e_0}{2}[\mathcal{A}^{\mu}_0(\phi_x)+\mathcal{A}^{\mu}_0(\phi_y)]+i\frac{e_0}{2}n^{\mu}\frac{((x-y)[\mathcal{A}_0(\phi_x)-\mathcal{A}_0(\phi_y))]}{\phi_x-\phi_y}\bigg\}.
\end{split}
\end{equation}
Let us now consider the vacuum case, where we expect that the vacuum four-current density identically vanishes. This can indeed be seen because for any finite $\epsilon^{\mu}$, we have that
\begin{equation}
\tilde{K}_0^{\mu}(x+\epsilon,x|0)+\tilde{K}_0^{\mu}(x,x+\epsilon|0)=-\frac{i}{8\pi^2}\int_0^{\infty}\frac{ds}{s^3}\,e^{-ism_0^2-i\epsilon^2/4s}(\epsilon^{\mu}-\epsilon^{\mu})=0.
\end{equation}
Thus, we are led to evaluate the quantity
\begin{equation}
\begin{split}
&\tilde{K}_0^{\mu}(x+\epsilon,x|\mathcal{A}_0)-\tilde{K}_0^{\mu}(x+\epsilon,x|0)+\tilde{K}_0^{\mu}(x,x+\epsilon|\mathcal{A}_0)-\tilde{K}_0^{\mu}(x,x+\epsilon|0)\\
&\quad=\frac{1}{4\pi^2}\int_0^{\infty}\frac{ds}{s^2}\,e^{-is\tilde{m}_0^2(\phi_x+\epsilon_-,\phi_x)-i\epsilon^2/4s}\bigg\{\mathcal{D}_1^{\mu}(x+\epsilon,x|\mathcal{A}_0)-i\frac{\epsilon^{\mu}}{2s}+i\frac{e_0}{2}[\mathcal{A}^{\mu}_0(\phi_x+\epsilon_-)+\mathcal{A}^{\mu}_0(\phi_x)]\\
&\qquad+i\frac{e_0}{2}n^{\mu}\frac{(\epsilon(\mathcal{A}_0(\phi_x+\epsilon_-)-\mathcal{A}_0(\phi_x))}{\epsilon_-}\bigg\}+i\frac{\epsilon^{\mu}}{8\pi^2}\int_0^{\infty}\frac{ds}{s^3}\,e^{-ism_0^2-i\epsilon^2/4s}\\
&\quad+\frac{1}{4\pi^2}\int_0^{\infty}\frac{ds}{s^2}\,e^{-is\tilde{m}_0^2(\phi_x,\phi_x+\epsilon_-)-i\epsilon^2/4s}\bigg\{\mathcal{D}_1^{\mu}(x,x+\epsilon|\mathcal{A}_0)+i\frac{\epsilon^{\mu}}{2s}+i\frac{e_0}{2}[\mathcal{A}^{\mu}_0(\phi_x)+\mathcal{A}^{\mu}_0(\phi_x+\epsilon_-)]\\
&\qquad+i\frac{e_0}{2}n^{\mu}\frac{(\epsilon(\mathcal{A}_0(\phi_x)-\mathcal{A}_0(\phi_x+\epsilon_-))}{\epsilon_-}\bigg\}-i\frac{\epsilon^{\mu}}{8\pi^2}\int_0^{\infty}\frac{ds}{s^3}\,e^{-ism_0^2-i\epsilon^2/4s},
\end{split}
\end{equation}
where $\epsilon_-=(n\epsilon)$. Now, we expand all functions up to the second-order terms in $\epsilon^{\mu}$. We need the expansions:
\begin{align}
&\tilde{m}_0^2(\phi_x+\epsilon_-,\phi_x)=\tilde{m}_0^2(\phi_x,\phi_x+\epsilon_-)\approx m_0^2-e_0^2\mathcal{A}_0^{\prime\, 2}(\phi_x)\frac{\epsilon_-^2}{12},\\
\begin{split}
&\mathcal{D}_1^{\mu}(x+\epsilon,x|\mathcal{A}_0)+i\frac{e_0}{2}[\mathcal{A}^{\mu}_0(\phi_x+\epsilon_-)+\mathcal{A}^{\mu}_0(\phi_x)]\approx-ie_0(\epsilon\mathcal{A}'(\phi_x))\frac{n^{\mu}}{2}+ie_0\mathcal{A}^{\prime\prime\,\mu}(\phi_x)\frac{\epsilon_-^2}{12}\\
&\quad-ie_0(\epsilon\mathcal{A}^{\prime\prime}(\phi_x))\frac{\epsilon_-}{3}n^{\mu},
\end{split}\\
\begin{split}
&\mathcal{D}_1^{\mu}(x,x+\epsilon|\mathcal{A}_0)+i\frac{e_0}{2}[\mathcal{A}^{\mu}_0(\phi_x+\epsilon_-)+\mathcal{A}^{\mu}_0(\phi_x)]\approx ie_0(\epsilon\mathcal{A}'(\phi_x))\frac{n^{\mu}}{2}+ie_0\mathcal{A}^{\prime\prime\,\mu}(\phi_x)\frac{\epsilon_-^2}{12}\\
&\quad+ie_0(\epsilon\mathcal{A}^{\prime\prime}(\phi_x))\frac{\epsilon_-}{6}n^{\mu}.
\end{split}
\end{align}
We observe that the convergence of the integrals for $s\to\infty$ is always ensured by the pre-exponential function. Moreover, the pre-exponential function is at least linear in $\epsilon^{\mu}$. Therefore, up to quadratic terms in $\epsilon^{\mu}$ we can approximate $\tilde{m}_0^2(\phi_x+\epsilon_-,\phi_x)=\tilde{m}_0^2(\phi_x,\phi_x+\epsilon_-)\approx m_0^2$ and we obtain
\begin{equation}
\begin{split}
&\tilde{K}_0^{\mu}(x+\epsilon,x|\mathcal{A}_0)-\tilde{K}_0^{\mu}(x+\epsilon,x|0)+\tilde{K}_0^{\mu}(x,x+\epsilon|\mathcal{A}_0)-\tilde{K}_0^{\mu}(x,x+\epsilon|0)\\
&\quad\approx\frac{1}{4\pi^2}\left[ie_0\mathcal{A}_0^{\prime\prime\,\mu}(\phi_x)\frac{\epsilon_-^2}{6}-ie_0n^{\mu}(\epsilon\mathcal{A}_0^{\prime\prime}(\phi_x))\frac{\epsilon_-}{6}\right]\int_0^{\infty}\frac{ds}{s^2}\,e^{-ism_0^2-i\epsilon^2/4s}.
\end{split}
\end{equation}
Now, by using the asymptotic expressions of the Hankel and the modified Bessel functions, one finds that the integral is approximately given by $-4i/(\epsilon^2-i0)$. Thus, in the limit $\epsilon^2\to 0$, the quantity $T^{(1),\mu}_0(x;\epsilon)$ is given by
\begin{equation}
T^{(1),\mu}_0(x;\epsilon)=\frac{e_0^2}{12\pi^2}\mathcal{F}^{\,\prime\,\mu}_{0\;\;\;\lambda}(\phi_x)n_{\rho}\frac{\epsilon^{\lambda}\epsilon^{\rho}}{\epsilon^2-i0},
\end{equation}
where $\mathcal{F}_0^{\mu\nu}(\phi_x)=n^{\mu}\mathcal{A}_0^{\prime\,\nu}(\phi_x)-n^{\nu}\mathcal{A}_0^{\prime\,\mu}(\phi_x)$ is the plane-wave electromagnetic field tensor. By finally taking the covariant limit for $\epsilon^{\mu}\to 0$, i.e., by replacing $\epsilon^{\lambda}\epsilon^{\rho}/(\epsilon^2-i0)\to \eta^{\lambda\rho}/4$, we conclude that $T^{\mu}_0(x)=0$. We point out how it has been important during the whole procedure not to take the limit $\epsilon^{\mu}\to 0$ (and then set $\epsilon^2=0$) in the term proportional to $\epsilon^2/s$ in the exponential function, the reason being that the quantity $s$ in the denominator assumes arbitrarily small values in the integration region.

The ambiguous integral representation of $T^{(1),\mu}_0(x)$ was obtained in Ref. \cite{Di_Piazza_2022_b} also by starting from the polarization operator in vacuum after noticing that the integrand was linear in the external field (see Fig. \ref{Tadpole_Pert_no_e-line}). Here, we would like to show how using dimensional regularization one can again conclude that $T^{(1),\mu}_0(x)$ vanishes. By applying Feynman rules in vacuum to the diagram in Fig. \ref{Tadpole_Pert_no_e-line}, we have that
\begin{equation}
\label{T0_PW}
T_0^{(1),\mu}(x)=-\int d^4y\,\text{Tr}(-ie_0\gamma^{\mu}iG^{(0)}_0(x-y|0)(-ie_0)\hat{\mathcal{A}}_0(\phi_y)iG^{(0)}_0(y-x|0)),
\end{equation}
where $G^{(0)}_0(x-y|0)$ is the tree-level, free electron propagator (see Eq. (\ref{G_0})). By writing the plane-wave four-vector potential as
\begin{equation}
\label{A_F}
\mathcal{A}_0^{\mu}(\phi_x)=\int\frac{d\omega}{2\pi}e^{-i\omega\phi_x}\check{\mathcal{A}}_0^{\mu}(\omega)=\int\frac{d\omega}{2\pi}e^{-i(kx)}\check{\mathcal{A}}_0^{\mu}(\omega),
\end{equation}
where $k^{\mu}=\omega n^{\mu}$, we obtain
\begin{equation}
\begin{split}
T_0^{(1),\mu}(x)&=-e_0^2\int \frac{d^4p}{(2\pi)^4}\int\frac{d\omega}{2\pi}\frac{e^{-i(kx)}}{p^2-m_0^2+i0}\frac{\text{Tr}(\gamma^{\mu}(\hat{p}+m_0)\hat{\check{\mathcal{A}}}_0(\omega)(\hat{p}-\hat{k}+m_0))}{(p-k)^2-m_0^2+i0}\\
&=-4e_0^2\int \frac{d^4p}{(2\pi)^4}\int\frac{d\omega}{2\pi}\frac{e^{-i(kx)}}{p^2-m_0^2+i0}\frac{(p\check{\mathcal{A}}_0(\omega))(2p^{\mu}-k^{\mu})-\check{\mathcal{A}}_0^{\mu}(\omega)(p^2-m_0^2-(kp))}{(p-k)^2-m_0^2+i0}.
\end{split}
\end{equation}
We now use the standard formula
\begin{equation}
\frac{1}{AB}=\int_0^1du\frac{1}{[A+u(B-A)]^2}
\end{equation}
and obtain
\begin{equation}
T_0^{(1),\mu}(x)=-4e_0^2\int\frac{d\omega}{2\pi}e^{-i(kx)}\int_0^1du\int \frac{d^4p}{(2\pi)^4}\frac{(p\check{\mathcal{A}}_0(\omega))(2p^{\mu}-k^{\mu})-\check{\mathcal{A}}_0^{\mu}(\omega)(p^2-m_0^2-(kp))}{(p^2-m_0^2-2u(kp)+i0)^2}.
\end{equation}
At this point, unlike in Ref. \cite{Di_Piazza_2022_b} we use dimensional regularization to manipulate the divergent integral in the four-momentum $p^{\mu}$. By calling $T_0^{(1),\mu}(x;d)$ the corresponding quantity in an arbitrary dimension $d$ and by introducing a mass scale parameter $\mu$, we have that
\begin{equation}
\begin{split}
T_0^{(1),\mu}(x;d)&=-4e_0^2\mu^{4-d}\int\frac{d\omega}{2\pi}e^{-i(kx)}\\
&\quad\times\int_0^1du\int \frac{d^dp}{(2\pi)^d}\frac{(p\check{\mathcal{A}}_0(\omega))(2p^{\mu}-k^{\mu})-\check{\mathcal{A}}_0^{\mu}(\omega)(p^2-m_0^2-(kp))}{(p^2-m_0^2-2u(kp)+i0)^2}.
\end{split}
\end{equation}
By performing the shift $p^{\mu}\to p^{\mu}+u k^{\mu}$, we obtain that the integrand does not depend on $u$ anymore and 
\begin{equation}
T_0^{(1),\mu}(x;d)=-4e_0^2\mu^{4-d}\int\frac{d\omega}{2\pi}e^{-i(kx)}\int \frac{d^dp}{(2\pi)^d}\frac{2(p\check{\mathcal{A}}_0(\omega))p^{\mu}-\check{\mathcal{A}}_0^{\mu}(\omega)(p^2-m_0^2)}{(p^2-m_0^2+i0)^2},
\end{equation}
where we have used the symmetry properties of the integrand. Finally, by using the standard integrals \cite{Peskin_b_1995}, we have
\begin{equation}
\begin{split}
T_0^{(1),\mu}(x;d)&=-4e_0^2\mu^{4-d}\int\frac{d\omega}{2\pi}e^{-i(kx)}\left[2\check{\mathcal{A}}_{0,\nu}(\omega)\frac{i\pi^{d/2}}{(2\pi)^d}\frac{\Gamma(1-d/2)}{2}\frac{-(m_0^2-i0)\eta^{\mu\nu}}{(m_0^2-i0)^{2-d/2}}\right.\\
&\quad\left.-\check{\mathcal{A}}_0^{\mu}(\omega)\frac{i\pi^{d/2}}{(2\pi)^d}\frac{-\Gamma(1-d/2)}{(m_0^2-i0)^{1-d/2}}\right]=0,
\end{split}
\end{equation}
and then we conclude that $T_0^{(1),\mu}(x)=\lim_{d\to 4}T_0^{(1),\mu}(x;d)=0$ as before.


%


\end{document}